\documentstyle[12pt,epsfig]{article}
\begin{document}
\parskip 7pt plus 2pt

\thispagestyle{empty}
\begin{flushright}
{CERN-TH/01-121}
\end{flushright}
\vspace*{1cm}  
\begin{center}
{\large{\bf The Cannonball Model of Gamma Ray Bursts:\\
high-energy neutrinos and $\gamma$-rays}}\\
\vspace{1cm}
Arnon Dar$^{\rm a}$ and
A. De R\'ujula$^{\rm b}$

\vspace*{.5cm}
$^{\rm a)}$ Physics Department and Space Research Institute, 
Technion, Haifa 32000, Israel\\
$^{\rm b)}$ Theory Division, CERN,
  1211 Geneva 23, Switzerland
\end{center}
\vspace{.3cm}
\begin{abstract}
\noindent
Recent observations suggest that $\gamma$-ray bursts (GRBs) and their
afterglows are produced by jets of highly relativistic cannonballs (CBs),
emitted in supernova (SN) explosions. The CBs,
reheated by their collision with the shell, emit radiation that is
collimated along their direction of motion and Doppler-boosted to the typical
few-hundred keV energy of the GRB. Accompanying the GRB, there should be
an intense burst of neutrinos of a few hundreds of GeV energy, made by
the decay of charged pions produced in the collisions of the CBs
with the SN shell . The neutrino beam carries almost all of the emitted
energy, but is much narrower than the GRB beam and should only be detected
in coincidence with the small fraction of GRBs whose CBs are moving very
close to the line of sight. The neutral pions made in the transparent
outskirts of the SN shell decay into energetic $\gamma$-rays (EGRs) of
energy of ${\cal{O}}$(100) GeV. The EGR beam, whose energy fluence is
comparable to that of the companion GRB, is as wide as the GRB beam and
should be observable, in coincidence with GRBs, with existing or planned
detectors.  We derive in detail these predictions of the CB model. 

\end{abstract}

\vspace{2.2cm}
\begin{flushleft}
{CERN-TH/01-121}\\
May  2001
\end{flushleft}
\newpage

\section{Introduction}

For a third of a century, gamma ray bursts (GRBs) have constituted a great
astrophysical mystery. Their origin is still an unresolved enigma, in
spite of recent remarkable observations in the field: the discovery of GRB
afterglows \cite{Costa, vP}, the discovery \cite{Galama} of the
association of GRBs with supernovae (SNe), and the measurements of the
redshifts \cite{Metz} of their host galaxies. The current generally
accepted view is that GRBs are generated by synchrotron emission from
relativistically expanding fireballs, or firecones, produced by collapses
or mergers of compact stars \cite{PG}, by failed supernovae or collapsars
\cite{WM}, or by hypernova explosions \cite{Pacz}. It was further
suggested that these highly relativistic fireballs produce large fluxes of
very high energy neutrinos in coincidence with the GRBs \cite{AHH}.  But
various observations suggest that most GRBs are produced by highly
collimated superluminal jets and not by relativistically expanding
fireballs \cite{SD, Dar98, DP, DD2000a}.

In a recent series of papers \cite{DD2000a, DD2000b, DD2001}
we have outlined a cannonball (CB)
model of GRBs which, we contend, is capable of describing the GRB
phenomenology, and results in interesting predictions.
The CB model is based on the following analogies, 
hypothesis and explicit calculations:

{\it Jets in astrophysics.}
Astrophysical systems, such as quasars and microquasars,
in which periods of intense accretion into a massive object occur, emit
highly collimated jets of plasma. The Lorentz factor 
$\rm \gamma\equiv 1/\sqrt{1-v^2/c^2}$ of these jets ranges from 
mildly relativistic: $\gamma\sim 2.55$ for PSR 1915+13 \cite{MR}, to
quite relativistic: $\gamma={\cal{O}}\,(10)$ for typical quasars \cite{Ghis},
and even to highly relativistic: $\gamma\sim 10^3$
for PKS 0405$-$385 \cite{Ked}. 
These jets are not continuous streams of matter, but consist
of individual blobs, or ``cannonballs''. The mechanism producing
these surprisingly energetic and collimated emissions is not
understood, but it seems to operate pervasively in nature.
We assume the CBs to be composed of ordinary ``baryonic''
matter (as opposed to $\rm e^+\, e^-$ pairs), as is the case in the microaquasar
SS 433, from which Ly$_\alpha$ and metal K$_\alpha$ lines have been detected
\cite{Marg, Kot}.

{\it The GRB/SN association.}
The original observation of a spatial and a temporal coincidence between GRB 
980425 and the relatively close-by  supernova SN 1998bw (redshift $\rm 
z=0.0085$), that suggested a physical association \cite{Galama},
has developed into a much more convincing case for the claim
\cite{DD2000a} that many, perhaps {\it all,} of the long-duration GRBs are
associated with SNe. Indeed, of the dozen and a half GRBs whose redshift
is known, the nearest six that have redshifts $\rm z<1$ show in their
afterglow an additive ``bump'', with the time dependence and spectrum of a
SN akin to 1998bw properly corrected \cite{DD2000a}, \cite{DDD} 
for the different redshift values and galactic extinction
\cite{DDD}: GRB 990228 \cite{Darz}, \cite{Reic}, \cite{Galamab}; GRB
970508 \cite{Sok}; GRB 980703 \cite{Holl}; GRB 990712 \cite{Hj},
\cite{Sahu};  GRB 991208 \cite{CT}; GRB 000418 \cite{DD2000a}.
  
In all the other cases with larger redshifts  there is one or
more good reasons for such a bump not to have been seen: no 
observations of the afterglow at late time are available, the expected bump
is below the sensitivity of the late time observations, 
the spectrum of SN 1998bw is not known at the frequencies
required to extrapolate its light curve to a much higher $\rm z$.  Thus,
observationally, seven out of sixteen ---and
perhaps all--- of the GRBs of known redshift have a SN associated with
them. The energy supply in a SN event similar to SN 1998bw is too small to
accommodate the fluence of cosmological GRBs, unless their $\gamma$-rays
are highly beamed. SN 1998bw is a peculiar supernova, but that may be due
to its being observed close to the axis of its GRB emission. It is not out
of the question that a good fraction --perhaps {\it all}-- of the
core-collapse SNe be associated with GRBs. To make the total cosmic rate
of GRBs and SN compatible, this nearly one-to-one GRB/SN association would
require beaming into a solid angle that is a fraction $\rm f\sim 2\times
10^{-6}$ of $4\pi$ \cite{DD2000a}. The CB model, for the emission from CBs
moving with $\gamma\sim 10^3$, implies precisely that beaming factor (the
numerical details are reproduced in Section 9). 

{\it The GRB engine.}
We assume a core-collapse SN event not to result only in the
formation of a central compact object and the expulsion of
a supernova shell (SNS). A fraction of the parent star's
material, external to the newly-born compact object,
should fall back in a time very roughly of the
order of one day \cite{DD2000a} and, given the considerable
specific angular momentum of stars,
it should settle into an accretion disk and/or torus
around the compact object\footnote{We choose to base
our conjectures on analogies with known processes, as opposed
to computer simulations. The latter do not yet realistically include rotation,
magnetic fields, the transport of angular momentum... Most
noticeably, they do not produce SN explosions.}.
The subsequent sudden episodes
of accretion ---occurring with a time sequence that we cannot predict---
result in the emission of CBs. These emissions last till
the reservoir of accreting matter is exhausted.
The emitted CBs initially expand in the SN rest-system at a speed 
$\rm\beta\,c/\gamma$, with $\rm\beta\,c$ presumably of the same order or 
smaller than the speed of sound in a relativistic plasma ($\beta=1/\sqrt{3}$).
The solid angle a CB subtends is so small that presumably
successive CBs do not hit the same point of the outgoing SNS,
as they catch up with it. These considerations 
are illustrated in Fig.(\ref{model}).

{\it The GRB.}
From this point onwards, the CB model is not based on analogies or
assumptions, but on processes whose outcome can be approximately 
worked out in an explicit manner. The violent collision of the CB
with the SNS heats the CB (which is not transparent at this point
to $\gamma$'s from $\pi^0$ decays) to a temperature that, by
the time the CB reaches the transparent
outskirts of the SNS, is $\sim 150$ eV,
further decreasing as the CB travels \cite{DD2000b}.
The resulting CB surface radiation, Doppler-shifted in energy 
and forward-collimated by the CB's fast motion, gives rise to an individual
pulse in a GRB, as illustrated in Fig.(\ref{model}).
The GRB light curve is an ensemble of such pulses,
often overlapping one another.
The energies of the individual GRB $\gamma$-rays, as well as
their typical total fluences, indicate CB Lorentz factors of 
${\cal{O}}$(10$^3$), as the SN/GRB association does \cite{DD2000b}.
The GRB properties most relevant to the current investigation are
reviewed in Section 5.

{\it The GRB's afterglow.}  The CBs, after they exit the SNS,
cool down by bremsstrahlung and radiate by this process,
by inverse Compton scattering, and by 
synchrotron radiation of their electrons on their enclosed magnetic field,
much as the plasmoids emitted by quasars and microquasars do
\cite{DD2000a}.  The CB model provides an excellent detailed 
description of optical afterglows \cite{DDD}.
The early afterglow spectrum and light curve are complicated
by the fact that, about a day after the GRB emission, CBs cool down to
a temperature at which $\rm e$--$\rm p$ recombination into $\rm H$
takes place. This gives rise to Ly-$\alpha$ lines that the CB's motion
Doppler-shifts to (cosmologically redshifted) energies of order 
a few keV, an energy domain that, interestingly,
coincides with that of the $\rm Fe$ lines that an object at rest would emit.
Recombination also gives rise to a multiband-flare in the afterglow.
These CB-model's expectations are in good agreement with incipient
data on X-ray lines and flares  \cite{DD2001}.

In all of the above considerations we have 
exploited the GRB/SN association to conclude that GRBs,
at least the long duration ones, are associated with core-collapse
SNe. This allows us to be very specific in our predictions
\cite{DD2000a, DD2000b, DD2001} concerning the collision
of the CBs with the SNS, and the consequent properties of the GRB
pulses (the density profile of  SNSs is known from observations;
the typical energy of CBs we can infer from the assumption
that the large peculiar velocities of neutron stars are
due to an imbalance between the momenta of the jets  of CBs they
emit as they are born \cite{DP}). But the sites of GRB emission
may not be only SNe. Any process of violent accretion,
such as a merger between neutron stars or other compact objects,
may result in the emission of CBs. If the latter encounter matter on
their way, such as circumstellar gas or a molecular cloud,
the processes leading to $\gamma$-ray emission would be similar
to the ones pertaining to a SN engine.

In this paper we address two other concrete predictions of
the Cannonball Model: the emission of neutrinos and of
energetic $\gamma$-rays (EGRs). Once again, to be specific,
we exploit our explicit model of GRBs emitted in core-collapse
SN events. The neutrinos are made by the chain decays of
charged pions, produced in the collisions of the CBs'
baryons with those of the SNS, as in Fig.(\ref{model}). 
The $\nu$ beam carries almost all of the emitted energy, but
is much narrower than the GRB beam and should only be
detected in coincidence with the small fraction of GRBs whose
CBs are moving extremely close to the line of sight.
The EGRs are made by the decay of neutral pions, but only from $\pi^0$
production close enough to the outskirts of the SNS for the $\gamma$-rays
not to be subsequently absorbed, see Fig.(\ref{model}).
The EGR beam, whose fluence is comparable to that of the GRB, is
as wide as the GRB beam and should be observable, in coincidence
with GRBs, with existing or planned detectors.
The EGR beam peaks at energies of tens of GeVs, while the
$\nu$ beam is about one order of magnitude more energetic.

\section{Times and energies}

Let $\rm \gamma=1/\sqrt{1-\beta^2}={E_{CB}/(M_{CB}c^2)}$ be 
the Lorentz factor
of a CB, which diminishes with time as the CB hits the SNS
and as it subsequently plows through the interstellar medium. Four
clocks ticking at different paces are relevant to a CB's history.
Let $\rm t_{SN}$ be the
local time in the SN rest system, $\rm t_{CB}$ the time in the CB's 
rest system, $\rm t_{Ob}$ the time   measured by
a nearby observer viewing the CB at an angle $\theta$
away from its direction of motion, and $\rm t$ the time
measured by an earthly observer viewing the CB at
the same angle, but from a ``cosmological'' distance 
(redshift $\rm z\neq 0$).
Let x be the distance traveled by the CB in the SN rest system.
The relations between the above quantities are:
\begin{eqnarray}
&&\rm
dt_{SN}=\gamma\,dt_{CB}=\rm{dx\over\beta\, c}\, ;
\nonumber \\
&&\rm
dt_{CB}\equiv \delta\,dt_{Ob}\, ;\nonumber\\ 
&&\rm
dt=(1+z)\,dt_{Ob}={1+z\over \gamma\,\delta}\;dt_{SN}\;,
\label{times}
\end{eqnarray}
where the Doppler factor $\delta$ is:
\begin{equation}
\rm
\delta\equiv\rm{1\over\gamma\,(1-\beta\cos\theta)}
\simeq\rm {2\,\gamma\over (1+\theta^2\gamma^2)}\; , 
\label{doppler} 
\end{equation}
and its approximate expression is valid for $\theta\ll 1$ and $\gamma\gg 1$,
the domain of interest here.
Notice that for large $\gamma$ and not large $\theta\gamma$,
there is an enormous ``relativistic aberration'':
$\rm dt\sim dt_{SN}/\gamma^2$, and the observer sees
a long CB story as a film in extremely fast motion.
 
The energy of the photons radiated by a CB
in its rest system, $\rm E^\gamma_{CB}$, their energy
in the direction $\theta$
in the local SN system, $\rm E^\gamma_{SN}$,  and the photon
energy $\rm E$ measured by a cosmologically distant observer,
are related by:
\begin{equation}
\rm E^\gamma_{CB}=   {E^\gamma_{SN}\over \delta}
\, ;\;\;\;\;\;E^\gamma_{SN}=(1+z)\,E\; ,
\label{energies}
\end{equation}
with $\delta$ as in Eq.(\ref{doppler}).

\section{Reference values of various parameters}

To be explicit we must scale our results to given values of the 
parameters of the CB model. In this section we introduce the
reference values that we adopt, which serve as benchmarks 
but imply no strong commitment to their particular choices.
These values are listed in Table I, for quick reference.

\begin{table}[h]
\hspace{2.5 cm}
\begin{tabular}{|l|c|c|c|}
\hline
\hline
$\;\;\;\;\;\;\;\;\;$Parameter   &Symbol &Value \\
\hline
SN-shell's mass     & $\rm M_S$             & $\rm 10\; M_\odot$ \\
SN-shell's radius   & $\rm R_S$              & $2.6\times 10^{14}$ cm \\
Outgoing Lorentz factor   & $\rm\gamma_{out}$  & $10^3$ \\
CB's energy   & $\rm E_{CB}$ & $10^{52}$ erg   \\
Initial $\rm v_{_T}/c$ of expansion & $\rm\beta_{in}$ & $1/(3\,\sqrt{3})$ \\
Final $\rm v_{_T}/c$ of expansion & $\rm\beta_{out}$ & $1/\sqrt{3}$ \\
\hline
Redshift   &z  &1   \\
CB's viewing angle  & $\theta$ & $\rm 10^{-3}$  \\
\hline
\hline
\end{tabular}
\caption{List of the ``reference'' values of various parameters. In the text a barred
parameter means its actual value divided by its reference value, so that,
for instance, $\rm \overline M_S=1/2$ means that the actual mass of the 
SN shell
is taken to be $\rm 5\,M_\odot$. Two parameters are not specific to our
model ($\rm z$ and $\theta$).}
\end{table}

Let ``jet'' stand for the ensemble of CBs emitted in one direction in a SN
event. If a momentum imbalance between the opposite-direction jets is
responsible for the large peculiar velocities of neutron stars,
${\rm v_{NS}\approx 450\pm 90~ km~s^{-1}}$ \cite{LL}, the
jet kinetic energy $\rm E_{jet}$ must be, as we shall assume for our GRB
engine, larger than $\rm M_{NS}\,v_{NS}\,c\sim 10^{52}$ erg,
for $\rm M_{NS}=1.4\,M_\odot$ \cite{DP}. 
We adopt a  value of $10^{53}$ ergs as the reference 
jet energy\footnote{The
jet-emitting process may be ``up--down'' symmetric to a  good
approximation, implying even bigger jet energies. 
 In the accretion of matter by black holes in quasars 
\cite{Celo, Ghi} and microquasars \cite{MR} the
efficiency for the conversion of gravitational binding energy into jet
energy appears to be surprisingly large. 
If in the production of CBs the central compact object
ingurgitates several solar masses, $\rm E_{jet}$ could be as large as
$\rm \sim M_\odot c^2\simeq 1.8\times 10^{54}$ erg.}. 
On average, GRBs have some five to ten significant pulses, so that the 
energy in a single CB may be 1/5 or 1/10 of $\rm E_{jet}$. We adopt 
$\rm E_{CB}=10^{52}$ erg as our reference value.
We denote with a
bar the actual value of a parameter in the units of its reference value
so that $\rm \overline{E}_{CB}$, for instance, means a given 
cannonball energy divided by $10^{52}$ erg.

Let $\rm \gamma_{in}$ be the Lorentz factor of a cannonball
as it is fired. We shall find $\rm\gamma_{in}={\cal{O}}(3\times 10^3)$
to be a ``typical'' value ($\rm\gamma_{in}$ is not an ``input'' parameter).
For this value and the reference CB energy, the CB's mass is
very small by stellar standards, and comparable to an Earth mass:
\begin{equation}
\rm M_{CB}\sim 0.6\, M_\otimes \;{3\times 10^3\over \gamma_{in}}\, .
\label{CBmass}
\end{equation} 
The baryonic number of the CB is:
\begin{equation}
\rm N_b\simeq {E_{CB}\over m_p\,c^2\,\gamma_{in}}\simeq 2.2\times 10^{51}\; 
\overline{E}_{CB}\, \left[{3\times 10^3\over\gamma_{in}}\right].
\label{NB}
\end{equation}
The collision of a CB with a SNS  is so violent ---at $\sim 1$ TeV
per nucleon--- that there is no doubt that, as it exits the shell,
the CBs' baryonic number resides in individual protons and neutrons.

We have assumed that, in a SN explosion, some of the material
outside the collapsing core is not expelled
as a SNS, but falls back onto the compact object. For vanishing angular momentum, the free-fall time of a test-particle from a distance $\rm R$
onto an object of mass ${\rm M}$ is
$\rm t_{fall} =\pi\,[R^3/(8\,G\,M)]^{1/2}$. For material falling
from a typical star radius ($\rm R_\star\sim 10^{12}$ cm) on an
object of mass $\rm M=1.4\;M_\odot$, $\rm t_{fall} \simeq 1$ day.
The fall-time is longer (except for material falling from the polar directions)
if the specific angular momentum is considerably
large, as it is in most stars. The fall-time is shorter for material
not falling from as far as the star's radius.
The estimate $\rm t_{fall} \simeq 1$ day is therefore a very rough one.
One day after core-collapse, the expelled SNS, traveling at
a velocity $\rm v_S \sim c/10$ \cite{Naka},
has moved to a distance:
\begin{equation}
\rm R_S=2.6 \times 10^{14} \;cm\;\left({t_{fall}\over 1\;d}\right)\;
\left({10\,v_S\over c}\right) .
\label{Rs}
\end{equation}
We adopt $\rm R_S=2.6 \times 10^{14}$ cm as our reference value.

For the Lorentz factor of the CBs as they exit the SNS, 
we adopt the value $\rm \gamma_{out}=10^3$, 
for the reasons discussed in the Introduction.
Let $\rm \beta_{in}\, c$ be the expansion velocity of a CB,
in its rest system, as it travels from the point of emission to
the point at which it reaches the SNS, and let $\rm \beta_{out}\, c$
be the corresponding value after the CB exits the SNS,
reheated by the collision. We expect these velocities to be comparable 
to the speed of sound in a relativistic plasma, $\rm c/\sqrt{3}$,
as observed in the initial expansion of the CBs
emitted by GRS 1915+105 \cite{MR}. As reference values, we
adopt those of Table I.

\section{The collision of a CB with the SNS}

\subsection{The shell's profile and transparency}

The density profile of the transparent outer layers of a SNS 
as a function
of the distance $\rm x$ to the SN centre can be inferred from the photometry,
spectroscopy and evolution of the SN emissions \cite{Naka}.
 The observations can be fit by a power law,
$\rm x^{-n}$, with $\rm n \sim 4\; to\, 8$. Our results 
for neutrino fluxes are not sensitive to this density profile 
and our results for GRB $\gamma$-rays \cite{DD2000b}
and for EGRs are only sensitive
to the outer region where the SN shell
becomes transparent. This implies that, for simplicity, we
can adopt at all $\rm x>R_S$ the density profile observed
in the shell's outer layers:
\begin{equation}
\rm \rho(x)=\rm\rho(R_S)\,\Theta(x-R_S)\,\left[{R_S\over x}\right]^n\, .
\label{profile}
\end{equation}
The SNS grammage still in front of a CB located at x is:
\begin{eqnarray}
&&\rm X_S(x)=\rm \int_x^\infty \, \rho(y)\,dy= X_{SNS}
\; \left[{R_S\over x}\right]^{n-1}\nonumber \\
&&\rm X_{SNS}\equiv\rm
{M_S\over 4\,\pi\, R_S^2}\simeq (2.35\times 10^4)\; 
{\overline{M}_S\over \overline{R}_S^2}
\;g \, cm^{-2}\; .
\label{SNgram}
\end{eqnarray}

For GRB photons in the MeV domain the attenuation length is similar, 
within a factor 2, in all elements from H to Fe, and it
is close to the attenuation length in a hydrogenic plasma. 
In the CB model, at a fixed time, the energy spectrum in a GRB
pulse is roughly thermal \cite{DD2000b}. The radiation length  
in the obscuring shell, averaged over a black body spectrum
of peak energy 1 MeV, is approximately: 
\begin{equation}
\rm X_{GRB}\simeq {m_p\over \sigma_{KN}(1\, MeV)}
\simeq  10\; g \, cm^{-2}\; ,
\label{XGRB}
\end{equation}
where $\rm \sigma_{KN}$ is the Klein-Nishina cross section.
 For EGRs the attenuation length in hydrogen 
in the 100 MeV to 100 GeV range, 
dominated by $\rm e^+\,e^-$ pair production, is \cite{Groom}:
\begin{equation}
\rm X_{EGR}\simeq 70 \; gr \, cm^{-2}\; .
\label{XEGR}
\end{equation}
The attenuation lengths of 
Eqs.(\ref{XGRB}) and (\ref{XEGR}) are all much smaller that the 
typical shell grammage of Eq.(\ref{SNgram}).

Equating $\rm X_S(x)$ and $\rm X_{GRB}$ 
and solving for x, one obtains the radial distance $\rm x_{GRB}^{tp}$ 
at which the SNS becomes (one radiation length) transparent to 
GRB photons.
For  our reference parameters, some representative results are: 
\begin{equation}
\rm
x_{GRB}^{tp}/R_S\simeq (3.7,\, 6.2,\, 21)\,\;\;\; for\; n=(8,\,6,\,4).
\label{xtp1}
\end{equation}
The corresponding
values for $\rm x_{EGR}^{tp}$, at a given $\rm n$, are shorter:
\begin{equation}
\rm
x_{EGR}^{tp}/R_S\simeq (2.3,\, 3.2,\, 6.9)\,\;\;\; for\; n=(8,\,6,\,4).
\label{xtp2}
\end{equation}
The GRB and EGR signals are emitted as the CB reaches
the transparent outskirts of the SNS. The neutrino signal
is emitted as soon as the CB starts colliding with the shell.
We discuss in detail in Section 10 the 
time profiles and relative timing of these signals.

\subsection{Kinematics of a CB's collision with a SN shell}

The radius of the expanding CBs, as they reach the SNS, is:
\begin{equation}   
\rm R_{CB}\sim R_S\,{\beta_{in}\over\gamma_{in}}
\simeq 1.7\times 10^{10}\;cm\;\overline\beta_{in}\,
\left[{3\times 10^3\over \gamma_{in} }\right]\,\overline R_S\, ,
\label{RCBatshell}
\end{equation}
In its collision with the shell, a CB sweeps up a ``target'' mass
\begin{equation} 
\rm M_T\sim\pi\,R_{CB}^2\,X_{SNS}=M_S\,
{\beta_{in}^2\over 4\,\gamma_{in}^2}
\simeq 3.5\times 10^{-3}\;M_\otimes\;\overline{\beta}_{in}\,
\left[{3\times 10^3\over \gamma_{in} }\right]^2\,\overline M_S\, ,
\label{MT}
\end{equation}
where $\rm X_{SNS}$ is the full column density of the shell,
as in Eq.(\ref{SNgram}). 

Seen from the reference system in which the CB is at rest
(and its shape, because of expansion, is roughly spherical)
the constituents of the SNS impinge onto the CB with
a Lorentz factor $\rm\gamma_{in}$. The average density
of a CB with the reference radius 
of Eq.(\ref{RCBatshell}) and the reference mass of Eq.(\ref{CBmass}) is
$\rm\rho\sim 1.8\times 10^{-4}$ gr cm$^{-3}$. The nucleon--nucleon 
interaction length
at that density is $\rm \lambda_{CB}=(N_A\,\sigma_{pp}^{TOT}\,\rho)^{-1}$
$\sim 2.2\times 10^5$ cm, with $\rm N_A$ Avogadro's number
and $\rm \sigma_{pp}^{TOT}\simeq 40$ mb the proton--proton
(or nucleon--nucleon) cross section at TeV beam energies.
The CB's radius of Eq.(\ref{RCBatshell}) is much
bigger than $\rm\lambda_{CB}$, implying that all nuclei in the region
of the shell swept up by the CB
interact. Approximately 1/3 of the energy in these
collisions results in photons from $\pi^0$ decay, which heat the CB
to a temperature in the keV domain \cite{DD2000a, DD2000b}.

Seen from the reference system in which the SNS is at rest
(or moving with a modestly relativistic velocity $\rm \sim c/10$) a
high-energy  nucleon in the CB ---suffering successive interactions in the 
dilute gas or plasma constituting the SNS---
loses roughly 2/3 of its energy to $\pi^\pm$
production. The density of the shell is of order 
$\rm\rho_S=M_S/(4\,\pi\,R_S^3)$ $\sim 10^{-10}$ gr/cm$^3$, for
our typical parameters. At that density, the nucleon-nucleon
interaction length is $\rm\lambda_S\sim 5 \times 10^{11}$ cm,
much less than the  $\cal{O}$($\rm R_S)$ shell's depth,
so that the shell's material is, in this sense, ``thick'': it acts as a beam 
dump. The decay length of a charged pion of energy $\rm E_\pi$ is
$\rm 5.6\times 10^5\,E_\pi/(100\, GeV)$ cm, much less
than its interaction length, which is comparable to that of nucleons.
Consequently, the beam dump is ``thin'' to $\pi$ decay and
roughly 2/3 of a CB's nucleon energy is
carried away by the neutrinos in $\pi\to \mu\,\nu$ decays
and in the subsequent $\mu$ decays.

The Lorentz factor of the CB after it has swept the SNS is simply the ratio 
of the total energy to the invariant mass of the outgoing object:
\begin{equation} 
\rm \gamma_{out}\simeq{E_{CB}/3\over
\sqrt{2\,M_T\,c^2\,E_{CB}/3+M^2_{CB}\,c^4}}\; ,
\label{gammaout1}
\end{equation}
where we have  used  $\rm E_{CB}\gg M_Tc^2$,
with $\rm M_T$ the target mass of Eq.(\ref{MT}). Substituting for
$\rm M_T$ and $\rm M_{CB}$ as functions of $\rm\gamma_{in}$
and $\rm\beta_{in}$, one obtains:
\begin{equation} 
\rm \gamma_{out} \simeq \gamma_{in} 
\;\sqrt{2\,E_{CB}\over 3\,\beta_{in}^2\,M_S\, c^2+18\,E_{CB}}
\label{gammaout}
\end{equation}
whose limiting values are:
\begin{eqnarray}
&& \rm \gamma_{out}\sim {\gamma_{in}\over 3}\;\;\;\; 
(for\; 6\,E_{CB}\gg \beta_{in}^2\,M_S\,c^2)\nonumber\\ 
&& \rm \gamma_{out}
\sim {\gamma_{in}\over 10\,\overline{\beta}_{in}}\,
\left[{\overline{E}_{CB}\over \overline{M}_S}\right]^{1\over 2}\;\;\;\;
(for\; 6\,E_{CB}\ll \beta_{in}^2\,M_S\,c^2)\; .
\label{gammaout2}
\end{eqnarray}

For our reference parameters, 
Eq.(\ref{gammaout}) implies that
$\rm\gamma_{in}\sim10\,\gamma_{out}$.
The very large ``typical'' values
of $\rm\gamma_{in}$, $\sim 3 \times 10^3$ or larger, 
as in Eqs.(\ref{gammaout2}),
imply that the fractional solid angle covered by a CB as it hits the
SNS is tiny: $\rm \beta_{in}^2/(4\,\gamma_{in}^2)\sim 10^{-9}$ or smaller,
for our reference $\rm\beta_{in}$.  This presumably makes it unlikely for 
consecutive CBs to hit precisely the same spot in the SNS:
CB--CB collisions and mergers may be the exception, rather than the rule,
and the collisional ``histories'' of successive CBs should be similar.

Let $\rm \sigma_{_T}\simeq 6.5\times 10^{-25}$ cm$^2$ be the
Thomson cross section, describing $\gamma$--$\rm e$
collisions at invariant masses
comparable or smaller than the electron mass.
The CB itself becomes transparent to the radiation it encloses
when it reaches a radius $\rm  R^{tp}_{CB}\simeq [3\,
M_{CB}\,\sigma_{_T}/(4\,\pi\,m_p)]^{1\over 2}$ $\sim 1.9\times 10^{13}$ cm,
for $\rm M_{CB}=0.6\,M_\otimes$. 
If the CB in its rest system, after the collision with the 
SNS,  is expanding at a transverse velocity $\rm \beta_{out}\, c$, the distance
away from the SN at which it becomes transparent is
$\rm\gamma_{out}\,R^{tp}_{CB}/\beta_{out}$, or
$\sim 3.3\times 10^{16}$ cm for our reference parameters. 
By then, the CB is well out of the SNS
and it has emitted from its expanding surface the radiation that constitutes
the GRB signal \cite{DD2000b}.

\subsection{Microscopic description of the collision}

The main point in outlining a microscopic picture of the collision
of a CB and a SNS, as we shall see, is to conclude that 
the details of such a picture are immaterial to the
estimate of the properties of GRBs and of their associated EGR-
and high-energy $\nu$ fluxes. But the discussion is important
in that it sets the basis for {\it how} to make these estimates.

Both the SNS and the CB are many $\rm pp$ interaction lengths long.
The number of such lengths in the SNS is:
\begin{equation}
\rm
N^{int}_{SNS}={M_S\;N_A\over\pi\,R_S^2}\,\sigma_{pp}^{TOT}
\simeq (5.6\times 10^2) \;\overline{M}_S\;[\overline{R}_S]^{-2}\, ,
\label{Nintshell}
\end{equation}
As it enters the shell at a distance $\rm x=R_S$ from the SN centre,
the number of $\rm pp$ interaction lengths in the CB is:
\begin{equation}
\rm
N^{int}_{CB}\sim{M_{CB}\;N_A\over\pi\,R_{CB}^2}\,\sigma_{pp}^{TOT}
\simeq (1.6\times 10^5)\;\overline{E}_{CB}\;
\left[{\gamma_{in}\over 3\times 10^3}\right]^2\;
[\overline\beta_{in}\,\overline{R}_S]^{-2}\; .
\label{NintCB1}
\end{equation}
At a later point in the crossing  of the SNS, e.g. at $\rm x=2\,R_S$,
 when the CB is moving at $\rm \gamma\sim \gamma_{out}$ 
and expanding at a speed $\rm\beta_{out}$, its number 
of interaction lengths is:
\begin{equation}
\rm
\widetilde N^{int}_{CB}\sim (5.1\times 10^2)
\;\overline{E}_{CB}\;[\overline\gamma_{in}]^{-1}\;
[\overline\gamma_{out}]^2\;
[\overline\beta_{out}\,\overline{R}_S]^{-2}\; .
\label{NintCB2}
\end{equation}
Thus, the number of $\rm pp$ interaction lengths in the CB is typically 
comparable to or bigger than the corresponding number in the SNS.

The simplest reference
system in which to visualize the collision is the centre-of-mass system (c.m.s.)
of two slabs of nuclei, one belonging to the CB, the other to the SNS, 
both one interaction length long.  Consider first the case in which the CB 
and the SNS  are an equal number
of nucleon--nucleon interaction lengths long.
In the approximation of constant densities, the slab--slab c.m.s.
coincides in this case with the overall c.m.s. of the CB and the
material that it hits in the SNS\footnote{It is easy to generalize
the argument that follows to non-constant densities, by using a
variable reference system in which the densities are instantaneously the 
same, but that requires an unjustified amount of effort, the final
results on the observable $\nu$ fluxes being the same. The
generalization to unequal numbers of interaction lengths,
we shall deal with explicitly.}.
In this system both the CB and the shell are spatially contracted
(relative to their respective rest systems) by the Lorentz factor
$\rm\sqrt{\gamma_{in}}/2$ at which their constituents are moving
towards each other. After the nucleons have interacted once,
their energy is degraded by an average factor $\rm f\sim 0.7$,
(the ``leading particle'' average-energy fraction observed
in high-energy nuclear collisions).
The nucleons of the leading slab, after the time required to
interact a few times with a few ``opposing'' slabs, come to rest
and are eventually turned back. Meanwhile fresh slabs are coming
in and suffering the same fate as the first. An increasingly
hot and dense pancake-shaped
region is formed, containing the nucleons that
have collided and the radiation initiated by $\gamma$'s from $\pi^0$ decay.
Because of its enclosed 
radiation pressure, this ``pancake''
eventually expands in its rest system at a speed
of $\cal{O}$($\rm c/\sqrt{3}$). When all the slabs of the CB
and of the SNS  are consumed, the resulting object is the
outgoing CB, at rest in this system.

In the case where the number of interaction lengths in the CB 
and the SNS are different, a similar description applies in the
system of reference in which the CB and shell densities
are the same, up to the moment in which the object with
the smaller number of interaction lengths is consumed.
This object is typically the SNS, as Eqs.(\ref{Nintshell})
to (\ref{NintCB2}) indicate.
At that point, we have a hot and dense pancake-shaped
object at rest, plus the fresh slabs from the CB's side of the collision
that are impinging on the disk without having interacted yet. 
As these fresh slabs hit, they set the disk in motion. The final
outgoing CB, now viewed from the 
local rest system of the parent SN, is moving with the Lorentz factor
$\rm\gamma_{out}$ of Eq.(\ref{gammaout}).

No doubt the previous description is oversimplified, for the violence
of the collision of the CB and the SNS presumably results in shocked and
turbulent motions. Moreover, the freshly expelled SNS is
probably not smooth, but also turbulent and inhomogeneous
on small scales. Fortunately, a detailed description is not required
for an estimate of the fluxes of $\nu$'s and EGRs.
 
\section{The GRB}

In this section we briefly review the properties\footnote{We use
the ``surface model'' of \cite{DD2000b}.} of GRBs, in the
CB model \cite{DD2000b}, that we need to establish comparisons
between the GRB itself, and its accompanying EGR and $\nu$ fluxes.

In its rest frame, the front surface of a CB is bombarded by the nuclei
of the SNS, which have an
energy $\rm m_p \,c^2\,\gamma\sim$ 1 TeV per nucleon, 
roughly 1/3 of which is converted into $\gamma$-rays 
(from $\pi^0\to\gamma\gamma$ decays)
within a nucleon attenuation length: 
\begin{equation}
\rm X_p={m_p\over \sigma^{TOT}_{pp}} \simeq 42 \; g \, cm^{-2}\; .
\label{Xp}
\end{equation}
These high energy photons initiate electromagnetic cascades that,  
in turn, convert their energy to thermal energy within the CB. 
The radiation length of high energy $\gamma$'s in hydrogenic plasma
 is $\rm X_{EGR}$,
given by Eq.(\ref{XEGR}). The energy of the electromagnetic cascade 
ends up as heat. The thermal photons, of energy $\rm E_\gamma \ll m_e\,c^2$,
have a radiation length:
\begin{equation}
\rm X_{T} \simeq {m_p\over \sigma_{_T}}
\simeq 2.6\;g\; cm^{-2}\, .
\label{XT}
\end{equation}

The thermal energy contained in a CB's front-surface layer 
of ``depth'' $\rm X_T$, continually supplied by the SNS 
incident nucleons, will be radiated away. It is reasonable to
expect that an equilibrium is established whereby, to
a fair approximation, the quasi-thermal emission rate from
the CB is in equilibrium with the fraction of energy deposited
by the CB's collision with the SN shell 
within this one-radiation deep layer. A
fraction $\rm X_T/X_p$ of the incoming protons interact
in the radiating layer, and a fraction $\rm X_T/X_{EGR}$ of the
energy of the $\gamma$s from $\pi^0$ production and decay is deposited 
in it. The total energy deposited by a single
SNS shell nucleon is $\rm \sim m_p\,c^2\,\gamma/3$.
Equating the energy deposition per unit time to that re-emitted
from the CB's surface as quasi-thermal
radiation, we obtain an instantaneous temperature:
\begin{equation}
\rm T(x)\simeq \left[{X_T^3\over X_p\, X_{EGR}}\,
{(n-1)\,c^3\, [\gamma(x)]^2 \over 
              6\,\sigma\,  x^{tp}_{GRB}}\right ]^{1/4}
              \left[ {x\over x^{tp}_{GRB}}\right]^{-n/4}\,, 
\label{Temperature}
\end{equation}
%
%
where $\sigma$ is the Stefan--Boltzmann constant,
$\rm\gamma(x)$ is a function evolving from $\rm \gamma_{in}$
to $\rm\gamma_{out}$, $\rm x^{tp}_{GRB}$ is as in Eqs.(\ref{xtp1}),
and $\rm n$ is the SNS density index of Eq.(\ref{profile}).
The temperature as the CB reaches the transparent region of the SNS,
$\rm T_{tp}\equiv T(x^{tp}_{GRB})$, is of order 0.15 keV, and is not very
sensitive to the parameters of the model \cite{DD2000b}, scaling roughly as:
\begin{equation}
\rm
T_{tp}\propto \left[{n-1\over R_S}\right]^{1/4}\;[\gamma_{out}]^{1/2}\, ,
\label{Ttp}
\end{equation}
since $\rm x_{tp}\propto R_S$ and, in the outer
regions of the SNS, $\rm \gamma(x)\sim \gamma_{out}$.

The time-width of a single-CB GRB pulse is roughly
characterized by a ``transparency time''
$\rm t_{GRB}$: the time elapsed
between the moment the CB enters the SNS and the time it
reaches its (one radiation length) transparent outer layer.
In the observer's frame, this time is:
\begin{equation}
\rm t_{GRB}\simeq  {1+z\over \gamma_{out}\,\delta}\;
{x^{tp}_{GRB} - R_S\over c}\; ,
\label{ttp}
\end{equation}
with $\rm x^{tp}_{GRB}$ as in Eq.(\ref{xtp1}). For our standard parameters
and a typical $\rm \theta\sim 3/\gamma_{out}$, $\rm t_{GRB}$ is 
$\sim 0.23,\,0.45,\,1.7$ s, for SNS indices $\rm n=8,\,6,\,4$, respectively.

The radius of the CB at the time
$\rm t=t_{GRB}$ is:
\begin{equation}
\rm R^{tp}_{CB}\sim R_{CB}+(x^{tp}_{GRB}-R_S)\,
{\beta_{out}\over\gamma_{out}}
\label{RCBtp}
\end{equation}
with $\rm R_{CB}$ as in Eq.(\ref{RCBatshell}). For times of
${\cal{O}}$$\rm (t_{GRB})$, the radius of the CB increases
approximately linearly with time.
In the CB's rest system, and at a fixed time, the
energy spectrum of the radiation emitted by the CB is an approximate
black-body spectrum, corrected for absorption in the SNS, and
emitted by a sphere whose surface grows as $\rm t^2$
and whose surface temperature decreases as $\rm [t_{GRB}/t]^{(n/2)}$,
as in Eq.(\ref{Temperature}). We have shown in \cite{DD2000b}
that this simple picture describes well the light curves
and energy spectra of GRB pulses. The predicted GRB
energy spectrum is the sum of thermal spectra with decreasing
temperatures. Its high energy tail is exponential, with a characteristic
temperature $\rm \sim T_{tp}$. At high energies, this is an underestimate
with respect to the observed GRB spectra, which decrease 
approximately as $\rm E^{-2}$.
We attribute this discrepancy to the naivet\'e of our thermal input 
spectrum: observations demonstrate that astrophysical plasmas
subject to a flux of high energy particles ---such as the CB in its rest system---
radiate a ``quasi-thermal'' spectrum, corrected at high energies
for such a power-law tail (clusters of galaxies are discussed in \cite{FF},
galaxy groups in \cite{YF} and SN remnants in \cite{SNR}).

To estimate the total energy radiated by a CB's heated surface
we must first compute the number $\rm N_p^{GRB}$
of SNS nucleons that provide
this energy, with the constraint that the radiation they eventually
produce be able to escape from the SNS. The naive estimate
$\rm N_p^{GRB}\simeq  \pi\;[R_{CB}^{tp}]^2\,X_{GRB}/ m_p$
turns out to be a very good approximation. Indeed, in
terms of $\rm n_p(x)$, the nucleon number density in the shell,
and $\rm X_{SNS}=X_S(R_S)$, the shell's total grammage of
Eq.(\ref{SNgram}),  $\rm N_p^{GRB}$  is:
\begin{eqnarray}
\rm N_p^{GRB} &=&\rm \int_{R_S}^\infty dx\; \pi\,\left[R_{CB}(x)\right]^2\,n_p(x)\;
Exp\left[-{m_p\over X_{GRB}}\,\int _X^\infty n_p(x')\,dx'\right]\nonumber\\ 
&=& \rm \pi\,{X_{GRB}\over m_p}\int_{R_S}^\infty\,dx\; \left[R_{CB}(x)\right]^2\;
{d\over dx} \, Exp\left[-{m_p\over X_{GRB}}\,\int_x^\infty n_p(x')\,dx'\right]
\nonumber\\ \rm
&\approx&\rm \pi\;[R_{CB}^{tp}]^2\,{X_{GRB}\over m_p}
\,\left(1-Exp[-X_S/X_{GRB}]\right)
\approx  \pi\;[R_{CB}^{tp}]^2\,{X_{GRB}\over m_p}\; ,
\label{NpGRB}
\end{eqnarray}
where we have approximated by a constant the radius
of the CB as it travels through the outer few GRB absorption
lengths. In the CB's rest frame, the total radiated energy is:
\begin{equation}
\rm E_{pulse}^{rest}\approx  
{X_{GRB}\,X_T^2\over X_{EGR}\,X_p}\,
{\pi\, 
[R_{CB}^{tp}]^2\,c^2\, \gamma_{out}
                      \over 3}\,.
\label{newenergy}  
\end{equation}
As an example, for our standard parameters and $\rm n=8$,
$\rm R_{CB}^{tp}\sim 4\times 10^{11}$ cm and
$\rm E_{pulse}^{rest}\sim 3\times 10^{45}$ erg.
The result scales roughly as $\rm R_S^2\,\beta_{out}^2/\gamma_{out}$
and could be one or two orders of magnitude smaller for
$\rm R_S$ and $\rm \beta_{out}$ somewhat below our reference values.

An observer at rest,
located at a known luminosity distance $\rm D_L(z)$ from the CB and
viewing it at an angle $\theta$ from its direction of motion, would measure
a  ``total'' (time- and energy-integrated) fluence per unit area:
\begin{equation}
\rm {df\over d\Omega}\simeq {1+z\over 4\,\pi\,D_L^2}
\,{E_{pulse}^{rest}}\;\left[
{2\,\gamma_{out} \over 1+\theta^2\,\gamma_{out}^2}\right]^3\; .
\label{dfdomega}
\end{equation}
In a critical ($\Omega=1$) Friedman universe the luminosity distance
is given by:
\begin{equation}
\rm D_L(z)= {(1+z)\, c\over  \,H_0}\;
\int_{1\over 1+z}^1\,{dx\over\sqrt{\Omega_\Lambda\,x^4+\Omega_M\,x}}\, ,
\label{lumdim}
\end{equation}
where $\rm H_0$ is Hubble's parameter, 
  ${\rm \Omega_M}$ and ${\rm \Omega_\Lambda=\Omega-\Omega_M}$, 
respectively, are the matter and vacuum
energy densities divided by the critical density 
(${\rm \rho_c=3\,H_0^2/8\pi\, G}$) and the radiation energy density has been 
neglected. In our explicit calculations we use
$\rm H_0=65$ km/(s Mpc), ${\rm \Omega_M}=0.3$ and 
${\rm \Omega_\Lambda}=0.7$, so that, for example, 
$\rm D_L(1)\simeq 7.12$ Gpc $\simeq 2.20\times 10^{28}$ cm. In 
Fig.(\ref{lumdis})
we show $\rm D_L(z)$ and $\rm [D_L(1)/D_L(z)]^2$
(the quantity to which we shall scale our results) for the quoted
cosmology and, for comparison, for the case $\rm\Omega_M=1$,
$\rm \Omega_\Lambda=0$, for which 
$\rm D_L(1)\simeq 5.41$ Gpc $\simeq 1.67\times 10^{28}$ cm.

\section{The sources of high energy particles}

\subsection{The origin of high energy $\nu$'s}

The nuclei or nucleons in the incoming CB are comoving with
it with a Lorentz factor $\rm\gamma_{in}$. Nuclei in the outgoing
CB have certainly been shattered into their constituent nucleons
by the violence of the collision between the CB and the shell.
They are comoving with the bulk Lorentz factor $\rm\gamma_{out}$.
On average, a high energy nucleon colliding at a large
centre-of-mass energy with a nucleon at rest
---or moving along the same direction---
exits the collision with a small transverse momentum and
a fraction $\rm f\sim 0.7$ of its original energy (for ultrarelativistic
particles this statement is independent of longitudinal Lorentz
boosts). The nucleons of the incoming CB must have their
Lorentz factor degraded from $\rm\gamma_{in}$ to
$\rm\gamma_{out}$. On average, this takes a number $\rm i$ of
high-energy collisions satisfying $\rm f^i=\gamma_{out}/\gamma_{in}$,
that is $\rm i \sim 3$, for $\rm\gamma_{in}\sim 3\,\gamma_{out}$,
or  $\rm i \sim 6$, for 
$\rm\gamma_{in}\sim 10\,\gamma_{out}$. 

Another way to reach a conclusion similar to the above 
is to view the interactions in the reference system
introduced in Section 4.3. There we saw that the bulk
of the CB's nucleons impinge, with a Lorentz factor $\rm\sqrt{\gamma_{in}}/2$,
on a ``pancake'', which is at rest, or moving towards them at a mildly
relativistic velocity $\rm \sim c/\sqrt{3}$. For $\rm\gamma_{in}=3\times 10^3$
($10^4$) the incoming energy of the CB's nucleons is $\rm E\sim 26$
GeV ($\sim 47$ GeV). Consider the number of interactions $\rm i$
necessary to bring these nucleons down to an energy ($\sim 5$ GeV)
below which the multiplicity of pion production on a stationary
target is no longer roughly 
constant (up to logarithmic corrections), but is suppressed by 
threshold effects. The argument of the previous paragraph now
yields $\rm i\sim 4$ (6) for $\rm\gamma_{in}=3\times 10^3$ ($10^4$).
One may be concerned with the fact that,
if the CB contains more interaction lengths than
the SNS's target funnel, the last of the CB's protons to interact with
the pancake (which is by then moving in the CB's direction) may
only suffer collisions at a centre-of-mass energy insufficient to
produce pions. To dissipate this concern, consider the very last
nucleon of the CB to suffer a collision with the ensemble of the CB
plus the swept-up mass, and go back to the system in which the
SN is at rest. In that system, the incoming nucleon and its target
have Lorentz factors $\rm\gamma_{in}$ and $\rm\gamma_{out}$,
respectively, which brings us back to the argument in the previous
paragraph. 

Another concern is that, if nucleons suffer only a few interactions with
a large centre-of-mass energy, the earlier estimate that 2/3 
of the CB's energy is lost to neutrinos may be grossly incorrect. But after
$\rm i$ interactions the fraction of the original energy of the CB's nucleons
that has been invested in pion-production is $\rm \alpha\simeq 1-f^i$,
two thirds of which ends up in neutrinos. For $\rm i=3$ (6), 
$\alpha\sim 66\%$ $(\sim 88\%)$, so that Eq.(\ref{gammaout}) is a fair
approximation.

Given the previous arguments,
we shall estimate the neutrino flux as that produced by
a nucleon beam, containing as many nucleons as the incoming CB, 
moving with a Lorentz factor $\rm\gamma_{in}$, and
interacting {\it thrice} on a nucleon or nuclear target. 
Because of ``Feynman scaling'', as we shall see, it does not
matter whether the target nucleons are at rest or receding
with a Lorentz factor $\rm\gamma_{out}$. It follows from the above 
discussion that this estimate of the neutrino flux is an underestimate.
However, in practice the incurred error is small, because
the charged pions produced in $\rm i  \ge 4$ interactions have
a relatively low energy. Since the detection efficiency of their decay
neutrinos is weighted, as we shall see, by two powers of energy,
the low-energy tail of the neutrino spectrum is immaterial.

\subsection{The origin of EGRs}

The SNS is only transparent to $\gamma$-rays in its
outer layer, some $\rm X_{EGR}=70$ g cm$^{-2}$ deep, Eq.(\ref{XEGR}).
That figure also corresponds to roughly two high-energy nucleon--nucleon 
interaction lengths, see Eq.(\ref{Xp}). 
The shell is many interaction lengths thick, so that by
the time the CB reaches the shell's $\gamma$-ray transparent outer layer,
it is already moving with $\rm\gamma\simeq\gamma_{out}$.
At that point the CB has reached a radius:
\begin{equation}
\rm
\widetilde R_{CB}\sim (x_{EGR}^{tp}-R_s)\;{\beta_{out}\over\gamma_{out}}\; .
\label{RCB'}
\end{equation}
For $\rm\beta_{out}=1/\sqrt{3}$, $\rm\gamma_{out}=10^3$
and the largest of the $\rm x_{EGR}^{tp}$ in Eq.(\ref{xtp2}),
$\rm \widetilde R_{CB}\sim 8.8 \times 10^{11}$ cm. 
With this radius and the reference mass of Eq.(\ref{CBmass})
the CB's grammage is 
$\rm M_{CB}/(\pi\,\widetilde R_{CB}^2)\sim 1500$  g cm$^{-2}$,
which is larger than that of the transparent outskirts of the shell,
$\rm X_{EGR}$.
We shall therefore compute the EGR flux as that originating
from the $\pi^0$'s made by the front of the CB as it interacts
with {\it all} the nucleons in the shell's transparent outer 
layer\footnote{We neglect the reinteractions with the shell's nucleons
that have already been struck one or more times and set
into forward motion, thereby slightly underestimating the
EGR flux.}. This total number, computed in the same way as
$\rm N_p^{GRB}$ in Eq.(\ref{NpGRB}), is:
\begin{equation}
\rm N_p^{EGR} \approx  \pi\;\widetilde R_{CB}^2\,{X_{EGR}\over m_p}\; ,
\label{Npfirst}
\end{equation}
whose numerical value is:
\begin{equation}
\rm N_p^{EGR} \simeq (1.4\times 10^{49})\,
\,\left[{x^{tp}_{EGR}-R_S\over 2.2\; R_S}\right]^2\,
\left[{\overline R_S \;
\overline\beta_{out}\over\overline\gamma_{out}}\right]^2\, ,
\label{Np}
\end{equation}
where the values of $\rm x^{tp}_{EGR}$ for various shell density 
indices $\rm n$ are those of Eq.(\ref{xtp2}) and
we have not made explicit the weak dependence on 
$\rm (R_S^2/M_S)$ to the power $\rm 1/(n-1)$.

The total energy of the EGR pulse generated by a CB, as seen by
a local observer at rest in the SN system, is:
\begin{equation}
\rm E_{EGR}\approx  {1\over 3}\;\pi\;\widetilde R_{CB}^2\,X_{EGR}
\; c^2\, \gamma_{out}\, .
\label{EEGR}
\end{equation}
To study the spectrum and angular collimation of the EGR flux,
as seen by a cosmologically distant observer,
we must recall the details of pion production and decay. 
 
\subsection{Pion production in nucleon--nucleon collisions}

The CB's baryon number, as it crosses the SNS, resides in protons or nuclei 
that are been broken into their constituents by relativistic collisions.
The same is the case for the funnel in the SNS that is swept up
by the CB. At TeV energies, the nucleon--nucleon, nucleon--nucleus
or nucleus--nucleus 
processes have different cross sections, but the properties
of the produced pions, {\it per colliding nucleon--nucleon pair,} are
very similar, and not significantly different for protons or neutrons.
Consequently, we can use for our considerations
the empirical information on pion (and kaon) production in proton--proton
collisions.

Bailly et al.~\cite{Bailly} reported on a study of inclusive 
charged pion production in the collisions of protons of energy 
360 GeV on a hydrogen target (c.m.s. energy $\rm \sqrt{s}\simeq 26$
GeV), as a function of ``Feynman x'':
\begin{equation}
\rm
x\equiv{2\,E_\pi^{cms}\over \sqrt{s}}\, .
\label{Feynman}
\end{equation}
The result, roughly the same within errors for $\pi^+$ and $\pi^-$
is:
\begin{equation}
\rm F_\pi(x)\equiv
{1\over\sigma_{pp}^{TOT}}\,
\int dp_{_T}^2\,{d\sigma^\pi\over dp_{_T}^2\, dx}\simeq 0.2\,{\pi\over x}
\,(1-x)^{3.6}\, ,
\label{xdistr}
\end{equation}
where $\rm p_{_T}$ is the pion transverse momentum.
Given the approximate isospin independence of the interactions,
the above result should also apply to the inclusive $\pi^0$
production.
The hypothesis of Feynman scaling, satisfied up to small
logarithmic corrections, is that Eq.(\ref{xdistr}) is independent
of energy. 

The x and $\rm p_{_T}^2$ dependences are observed to
factorize to a good approximation, and the $\rm d\sigma^\pi/dp_{_T}^2$
distribution
is roughly exponential, with average 
\begin{equation}
\rm \bar p_{_T}^\pi\sim 320 \;\, MeV.
\label{pt}
\end{equation}
The most precise data on transverse-momentum distributions
at the relevant c.m.s. energies were collected 
for $\pi^0$ production in pp interactions at the ISR collider \cite{Neu};
the measured single-photon yield resulted in 
$\rm \bar p_{_T}^\gamma\sim 160$ MeV, whence the result we adopt
for neutral or charged pions ($\rm p_{_T}^\pi\simeq 2\,  p_{_T}^\gamma$).
The double differential cross section for inclusive $\pi$ production
is therefore of the form:
\begin{eqnarray}
\rm 
{1\over\sigma_{pp}^{TOT}}\,
{d\sigma^\pi\over dp_{_T}^2\, dx}\simeq && \rm
F_\pi(x)\;G_\pi(p_{_T})\; , \nonumber \\
\rm
G_\pi(p_{_T})\simeq && \rm
{1\over 2\, \bar{p}_T^2}\,e^{-p_{_T}/\bar p_{_T}}\; ,
\label{pTxdistr}
\end{eqnarray}
with $\rm F_\pi(x)$ as in Eq.(\ref{xdistr}) and $\rm \bar p_{_T}$
as in Eq.(\ref{pt}).
For ultrarelativistic reaction products, Eq.(\ref{pTxdistr})
is invariant under longitudinal Lorentz boosts, so that,
with $\rm x\simeq {E_\pi^{lab}/ E_p^{lab}}$, it can be used
for proton interactions on a stationary target.
We shall use it for incident $\rm E_p$  up
to a few TeV.

\section{The flux of EGRs}

To compute the spectrum of outgoing photons {\it per nucleon--nucleon
collision}, we must convolute Eq.(\ref{pTxdistr}) with the distribution
of $\pi^0\to\gamma\gamma$ decay. For ultrarelativistic pions, 
the distribution of fractional photon
energies ($\rm w\equiv E_\gamma/E_\pi$)
is flat and limited by $\rm 0<w<1$. 
The $\gamma$ distribution
in $\rm y\equiv E_\gamma/E_p$, produced by the decay of
pions distributed as in Eq.(\ref{xdistr}), is:
\begin{equation}
\rm
F_\gamma(y)=2\;\int_0^1 F_\pi(x)\; {dx\over x}
\int_0^1 dw \;\delta\left(w-{y\over x}\right)\; ,
\label{xpigammas}
\end{equation}
where the prefactor is for the two $\gamma$'s per $\pi^0$ decay.
To a few per cent accuracy, the result of the convolution
can be fitted\footnote{An exponential fit is inadequate close
to the limit $\rm y=1$, but for $\rm y>1/2$ the
flux is negligible.} by:
\begin{eqnarray}
\rm
F_\gamma(y)&\simeq & \rm A_\gamma\,{1\over y}
\,e^{-b_\gamma y}\\ \rm
A_\gamma&\simeq & \rm 1.1\; ,\;\;\; b_\gamma\simeq 8\; .
\label{fgamma}
\end{eqnarray}

The $\gamma$-production double differential cross-section in $\rm y$ and 
$\rm p_{_T}^\gamma$ is of the same form as Eq.(\ref{pTxdistr}),
with a longitudinal factor $\rm F_\gamma(y)$ and a transverse factor
with $\rm \bar p_{_T}^\gamma\sim 160$ MeV. 
Let $\rm E_\gamma$ be the energy of a photon as it
reaches the Earth, cosmologically
red-shifted by a factor $\rm 1+z$, and let
$\rm E_p\simeq m_p\,c^2\,\gamma_{out}$ be the energy 
of the CB's nucleons,
 in the local rest system of their SN progenitor,
as they reach the outer part of the SNS. For the
small angles $\theta$ at which the $\gamma$-rays are forward-collimated
by the relativistic motion of the parent $\pi^0$'s, the photon-number
distribution in $\rm x_\gamma=E_\gamma/E_p$
and $\cos\theta$, per single nucleon--nucleon
collision, is:
\begin{eqnarray}
\rm
{dn_\gamma\over dx_\gamma\;d\cos\theta}&\simeq & \rm
B_\gamma\; (1+z)^2\;x_\gamma\; e^{-c_\gamma\,x_\gamma}\nonumber\\ \rm
B_\gamma&\simeq&\rm A_\gamma\;
\left[{m_p\,\gamma_{out}\over \bar p_{_T}^\gamma}\right]^2
\simeq (3.76\times 10^7)\,\left[{\gamma_{out}\over 10^3}\right]^2\nonumber\\ \rm
c_\gamma&= &\rm c_\gamma (z,\theta,\gamma_{out})\simeq \rm ( 1+z)\,
\left[b_\gamma+{m_p\,\gamma_{out}\,\theta\over\bar p_{_T}^\gamma}\right]\, .
\label{xtheta1}
\end{eqnarray}

Let $\rm dn_\gamma/d\Omega$ be the total (time-integrated)
number flux of EGR photons
per unit solid angle about the direction $\theta$ (relative to
the CBs' direction of motion)  at which they
are viewed from Earth. 
The photon number distribution  per incident CB is:
\begin{eqnarray}
\rm 
{dn_\gamma\over dx_\gamma\, d\Omega}&\sim& \rm
{N_p^{EGR}\,B_\gamma\over 2\,\pi\;D_L^2}\;(1+z)^4\;f_\gamma\nonumber\\
\rm f_\gamma&\equiv&\rm f_\gamma (z,\gamma_{out},\theta,x_\gamma)
\simeq
x_\gamma\; e^{-c_\gamma\,x_\gamma}\; ,
\label{photnum}
\end{eqnarray}
with $\rm N_p^{EGR}$ given by Eq.(\ref{Np}). Since a
typical GRB has an average of $\rm n_{CB}=5$ to 10 significant pulses,
the total flux of EGRs in coincidence with a GRB may be
an order of magnitude above that of Eq.(\ref{photnum}).
In Fig.(\ref{spectrax}a) we show $\rm f_\gamma$ as a function of
$\rm x_\gamma$ at various $\theta$;
for $\rm z=1$ and $\rm\gamma_{out}=10^3$. 
The average fractional EGR energy in the spectrum of Eq.(\ref{photnum})
is $\rm\bar x_\gamma=2/c_\gamma$, corresponding, at
$\rm z=1$ and for $\rm\gamma_{out}=10^3$, to average energies
$\rm\bar E_\gamma\sim 120$ GeV
for $\theta=0$, $\rm\bar E_\gamma\sim 70$ GeV for 
$\rm \theta=1/\gamma_{out}$,
and $\rm\bar E_\gamma\sim 40$ GeV for $\rm \theta=3/\gamma_{out}$, 
a more probable angle of detection \cite{DD2000a}.
Except at the highest of these energies and/or at redshifts
well above unity, the absorption of $\gamma$-rays on the infrared
background ---for which we have not explicitly
corrected Eq.(\ref{photnum})--- is negligible.

Roughly characterize the efficiency of a $\gamma$-ray detector
as a step function $\rm \Theta(E^\gamma-E^\gamma_{min})$.
The total flux above threshold, per incident CB, is then:
\begin{eqnarray}
\rm {dn^T_\gamma[x^\gamma_{min},\theta]\over d\Omega}&\sim&\rm
{dn^T_\gamma[0,0]\over d\Omega}\;
G_\gamma (z,\gamma_{out},\theta,x^\gamma_{min})\nonumber\\ \rm   
G_\gamma&\simeq&\rm \left[{(1+z)\,b_\gamma\over c_\gamma}\right]^2
\,(1+c_\gamma\,x^\gamma_{min})\,
e^{-c_\gamma\,x^\gamma_{min}}\nonumber\\ \rm
x^\gamma_{min}&\equiv&\rm {E^\gamma_{min}\over m_p\,\gamma_{out}}
\nonumber\\ \rm
{dn^T_\gamma[0,0]\over d\Omega}
&\simeq&\rm {1.1\times 10^8\over km^2}\,{N_p^{EGR}\over 1.4\;10^{49}}\,
\overline\gamma_{out}^2  
\left[{1+z\over 2}\right]^2\,\left[{D_L(1)\over D_L(z)}\right]^2\, ,
\label{photflux2}
\end{eqnarray}
where the scaling properties of $\rm N_p^{EGR}$ are those of Eq.(\ref{Np}).
In Fig.(\ref{Ggamma}a) and (\ref{Ggamma}b) 
we show $\rm G_\gamma$ as a function of
$\rm x^\gamma_{min}$ at various fixed $\theta$, and vice versa;
for $\rm z=1$ and $\rm\gamma_{out}=10^3$. The very large flux
$\rm dn^T_\gamma[0,0]/ d\Omega$ of Eq.(\ref{photflux2}) is seen
to be significantly reduced as soon as $\theta$ and/or $\rm x^\gamma_{min}$
depart from zero: the EGR flux is not as gigantic as it appears to be
at first sight.

\subsection{EGR versus GRB total energies}

An observer at a cosmological distance from a GRB source, if equipped
with a detector of sufficient angular coverage, would measure a total
energy per CB, in the MeV-range photons of a GRB pulse, of
$\rm E^z_{GRB}=\gamma_{out}\,E^{rest}_{pulse}/(1+z)$,
with $\rm E^{rest}_{pulse}$ given by Eq.(\ref{newenergy}).
In the multi-GeV range of the EGRs, the measurement would
result in $\rm E^z_{EGR}=E_{EGR}/(1+z)$,
with $\rm E^{EGR}$ given by Eq.(\ref{EEGR}).
The ratio of EGR to GRB total energies is:
\begin{eqnarray}
\rm {E^z_{EGR}\over E^z_{GRB}}  &\simeq& \rm r \;
{X_{EGR}^2\, X_p\over X_{GRB}\, X_T^2}\;
{1\over \gamma_{out}} \nonumber \\
\rm r &\equiv & \rm {R_{EGR}^2 \over R_{GRB}^2}
\simeq \left[{x_{EGR}^{tp}-R_S \over x_{GRB}^{tp}-R_S}\right]^2 \, .
\label{ratio}
\end{eqnarray}
The factor $\rm 1/\gamma_{out}$ may look surprising at first:
the total GRB energy is $\propto \gamma^2$, while the EGR
energy is $\propto \gamma$. In both cases, one factor of
$\gamma$ is associated with individual nucleon energies.
But GRB photons ``benefit'' from the energy associated with 
the bulk motion of the CB, which acts as a relativistic mirror
(target SNS particles at rest, if elastically and coherently
back-scattered by the much heavier 
CB, would recoil with a Lorentz factor $\gamma^2$, while
particles produced in individual $\rm pp$ collisions would
carry an energy scaling as $\gamma$).

Using the results of Eqs.(\ref{xtp1}) and (\ref{xtp2}) for 
the transparency distances, we obtain:
\begin{equation}
\rm {E^z_{EGR}\over E^z_{GRB}}  \simeq
(0.71,\,0.54,\,0.26)\;[\overline \gamma_{out}]^{-1}\;\;\;\;\;\;\;
for\; n=(8,\,6,\,4),
\label{ratio2}
\end{equation}
that is, the EGRs carry almost as much energy as the GRB.
Since the individual EGR photons have energies of
${\cal{O}}(100)$ GeV, as opposed to ${\cal{O}}(1/2)$ MeV
for GRB photons, the number of EGR photons is five or
six orders of magnitude below that in the associated GRB.

\section{The flux of high energy neutrinos}

The calculation of the $\nu_\mu$ flux 
produced in the collision of  a CB with the 
SNS is analogous to the calculation
of the photon flux. The $\bar\nu_\mu$ flux gives rise to a signal
of about 1/3 the size of that of the $\nu_\mu$ flux
(we neglect it, since we find it preferable to establish
a lower limit to the observational prospects).
The $\nu_\mu$'s are made in the chain reactions
$\rm p\,p\to \pi + ...$, $\pi^+ \to \mu^+\,\nu_\mu$; and 
$\pi^-\to\mu^-\,\bar{\nu}_\mu$, followed by 
$\mu^-\to e^-\,\nu_\mu\,\bar{\nu}_e$.
We have also estimated the contribution of $\rm K$
production and decay, which turns out to be negligible.

\subsection{Pion and muon decay}

We have shown in section 6.1 that, in order to estimate the
neutrino flux from the CB--SNS collision, it is adequate to consider
an average of $\rm i\sim3$ interactions of the incoming nucleons.
The leading outgoing particle in these
interactions has an average transverse momentum significantly
smaller than that of the produced mesons: it can be
neglected. The simplest way to compute $\nu$ fluxes is to
work out first the pion  distributions made in three
successive interactions of the incoming nucleons and then
convolute the result with the pion decay distributions. 
The pions made in the first interaction have the distribution of Eq.(\ref{xdistr}).
To compute the distribution of those made in the two successive
interactions, we must make convolutions, analogous to that in
Eq.(\ref{xpigammas}), with the distribution of the 
 exiting leading-nucleon longitudinal-momentum distribution.
For the latter, we use the measurements of Ref. \cite{BaBre}.
The result, $\rm F_\pi^{[3]}$, of summing the pion longitudinal distributions
as functions of $\rm x=E_\pi/E_p$ (with $\rm E_p$ the original
incoming nucleon's energy) can be simply parametrized,
to $\sim 15$\% accuracy, as:
\begin{equation}
\rm F_\pi^{^{[3]}}(x)\simeq [1+2.2\, (1-x)^4]\;F_\pi(x)\; ,
\label{Fpi3}
\end{equation}
with $\rm F_\pi(x)$ given by Eq.(\ref{xdistr}).

In the decay in flight
$\pi^-\to\mu^-\,\bar{\nu}_\mu$ of pions with $\rm E_\pi\gg m_\pi$, 
the distributions of fractional
neutrino and muon energies ($\rm x_\nu\equiv E_\nu/E_\pi$
and $\rm x_\mu\equiv E_\mu/E_\pi$) are flat and
limited by $\rm 0<x_\nu<x_{max}\equiv 1-m_\mu^2/m_\pi^2$
and $\rm 1-x_{max}<x_\mu <1$. The $\nu_\mu$ distribution
in $\rm y\equiv E_\nu/E_p$, produced by the decay of
pions distributed as in Eq.(\ref{xdistr}), is given by:
\begin{equation}
\rm
F_\nu(y)=\int_0^1 F_\pi^{^{[3]}}(x)\; {dx\over x}
\int_0^{x_{max}} {dx_\nu\over x_{max}} \;\delta\left(x_\nu-{y\over x}\right)\; ;
\label{xpinumu}
\end{equation}
the muon distribution in the decay $\pi^-\to\mu^-\,\bar{\nu}_\mu$ 
is analogous, with the proper change of integration limits. 
The $\nu_\mu$ distribution in $\rm z\equiv E_\nu/E_\mu$
in the decay of left-handed muons is $\rm 3\,z^2$ and, upon
neglect of $\rm m_e^2/ m_\mu^2$, it extends from 0 to 1.
We do not write here explicitly the double convolution, analogous
to Eq.(\ref{xpinumu}),  involved in the calculation of the y-distribution
in $\pi\to\mu\to\nu_\mu$ decay.

The mean $\rm x_\nu= E_\nu/E_\pi$ in $\pi\to\mu\nu$ decay is
$\rm \bar x_\nu=x_{max}/2\simeq 0.19$, while for the
muon $\bar x_\mu\simeq 0.81$.
The mean $\rm z= E_\nu/E_\mu$ in $\pi\to\nu_\mu\, ...$ decay
is $\rm \bar z=3/4$, so that the mean $\rm x_\nu$
in the $\pi\to\mu\to\nu_\mu$ decay chain is 
$\bar x'_\mu\simeq \bar x_\mu\,\bar z \simeq 0.6$.
The available rest energy in $\pi\to\mu\nu$ or $\rm \mu\to e\nu\nu$
decay is small relative to the mean transverse momentum
of the parent pion. This implies that the mean $\nu_\mu$
 transverse momentum in the $\pi\to\mu\nu$ chain is 
$\rm \bar x_\nu\,\bar p_{T}\simeq 60$ MeV, where we have used
Eq.(\ref{pt}). The corresponding result for the $\pi\to\mu\to\nu_\mu$
chain is $\rm \bar x'_\nu\,\bar p_{T}\simeq 190$ MeV.
We shall see that the muon detection sensitivity on Earth
is weighted by two powers of energy (one for the cross section,
one for the muon range), so that the $\pi\to\mu\to\nu_\mu$
process, which produces a harder $\nu_\mu$ beam, is harder.
Rather than giving results for a two-component distribution
($\pi\to\mu\nu$ and $\rm \mu\to e\nu\nu$)
we shall use a common transverse momentum:
\begin{equation}
\rm \bar p_{_T}^\nu=190\;\;MeV
\label{ptnu}
\end{equation}
for the overall $\nu_\mu$ beam. This results in a
small underestimate of the flux at a fixed angle.

The relative contribution of kaons to the $\nu_\mu$ flux
is suppressed with respect to that of pions for three reasons.
The $\rm K/\pi$ relative multiplicity is $\sim 1/5$; the
$\rm K\to \mu\nu$ branching ratio is $\sim 63$\%; and
the available energy in the decay is not negligible in comparison with
$\rm \bar p_{_T}^K$. All in all, the $\rm K$-decay contribution
to the $\nu_\mu$ flux at fixed angle is at the few per cent level:
we neglect it altogether.

The result of all this analysis is a longitudinal distribution
in $\rm y=E_\nu/E_p$ (with $\rm E_p$ the incoming nucleon's
energy) that can, to
a few per cent accuracy, be fitted by:
\begin{eqnarray}
\rm
F_\nu(y)&\simeq & \rm A_\nu\,{1\over y}
\,e^{-b_\nu y}\\ \rm
A_\nu&\simeq & \rm 3\; ,\;\;\; b_\nu\simeq 12\; .
\label{fnu}
\end{eqnarray}
The $\nu_\mu$-production double differential cross-section in $\rm y$ and 
$\rm p_{_T}^\nu$ ---describing the neutrino flux
generated in the beam dump per incident proton---
 is of the same form as Eq.(\ref{pTxdistr}),
with a longitudinal factor $\rm F_\nu(y)$ and a transverse factor
with $\rm \bar p_{_T}^\nu\sim 190$ MeV.

Let $\rm E_\nu$ be the cosmologically redshifted energy of a neutrino as it
reaches the Earth, and let
$\rm E_p\simeq m_p\,c^2\,\gamma_{in}$ be the energy 
of the CB's nucleons, in the local rest system of their SN progenitor, 
as they enter the SNS. In analogy with Eq.(\ref{xtheta1}) the $\nu_\mu$-number
distribution in $\rm x_\nu=E_\nu/E_p$
and $\cos\theta$, per single nucleon--nucleon collision, is:
\begin{eqnarray}
\rm
{dn_\nu\over dx_\nu\;d\cos\theta}&\simeq & \rm
B_\nu\; (1+z)^2\;x_\nu\; e^{-c_\nu\,x_\nu}\nonumber\\ \rm
B_\nu&\simeq&\rm A_\nu\;
\left[{m_p\,\gamma_{in}\over \bar p_{_T}^\nu}\right]^2\simeq
(6.0\times 10^7)\,\left[{\gamma_{in}\over 10^4}\right]^2\nonumber\\ \rm
c_\nu&= & \rm c_\nu (z,\theta,\gamma_{in})
\simeq  ( 1+z)\,
\left[b_\nu+{m_p\,\gamma_{in}\,\theta\over\bar p_{_T}^\nu}\right]\, .
\label{xtheta3}
\end{eqnarray}

Let $\rm dn_\nu/d\Omega$ be the time-integrated number of neutrinos
per unit solid angle about the direction $\theta$ (relative to
the CBs' direction of motion)  at which they
are viewed from Earth. In analogy with Eq.(\ref{photnum}),
the neutrino number distribution, per incident CB, is:
\begin{eqnarray}
\rm 
{dn_\nu\over dx_\nu\, d\Omega}&=& \rm
{N_b\,B_\nu\over 2\,\pi\;D_L^2}\;(1+z)^4\;f_\nu\nonumber\\ \rm
f_\nu&=&\rm f_\nu(z,\gamma_{in},\theta,x_\nu)\simeq
x_\nu\; e^{-c_\nu\,x_\nu}\; ,
\label{nunum}
\end{eqnarray}
with $\rm N_b$ the total baryon number of the CB,
given by Eq.(\ref{NB}). For a GRB with $\rm n_{CB}$ significant pulses,
the total number of neutrinos is $\rm n_{CB}$ times
larger than that of  Eq.(\ref{nunum}).

In Fig.(\ref{spectrax}b) we show $\rm f_\nu$ as a function of
$\rm x_\nu$ at various $\theta$;
for $\rm z=1$ and $\rm\gamma_{in}=10^4$. 
The average fractional $\nu$ energy in the spectrum of Eq.(\ref{nunum})
is $\rm\bar x_\nu=2/c_\nu$, corresponding, for the chosen $\rm z$
and $\rm\gamma_{in}$, to average energies
$\rm\bar E_\nu\sim 712$ GeV
for $\theta=0$, $\rm\bar E_\nu\sim 315$ GeV for 
$\rm \theta=1/10^3$,
and $\rm\bar E_\nu\sim 150$ GeV for $\rm \theta=3/10^3$.

Neutrino oscillations may reduce the flux of $\nu_\mu$s of
Eq.(\ref{nunum}) by as much as a factor of 2 (if they are
maximal) or even 3 (if they are ``bimaximal'').

\subsection{Muon production on Earth}

Muon neutrinos produced by a GRB can be detected by large-area
or large-volume detectors, in temporal and directional
coincidence with a GRB $\gamma$-ray signal. 
The detection technique typically involves
the ``upward-going'' muons, for which there is no ``atmospheric''
cosmic-ray background. 

A flux of neutrinos traversing rock or ice 
interacts with target nuclei N, producing muons
in the process $\rm \nu_\mu + N\to \mu + ...$ In the energy
range of interest here, the inclusive
muon cross-section per target nucleon is:
\begin{equation}
\rm
{d\sigma(E_\nu,E_\mu)\over dE_\mu}\simeq {\sigma_{_{CC}}\over E_\nu}\;
\theta(E_\nu-E_\mu);\;\;\;\;\;\;\;
\sigma_{_{CC}}\simeq 0.8\times 10^{-38}\;cm^2\;{E_\nu\over GeV}\; .
\end{equation}

The produced muons lose energy and ``range-out'' in matter
before they decay. At the energies of interest here,
the muon energy loss per unit distance x (in a material
of average atomic number and mass Z and A) can be
approximated by:
\begin{equation}
\rm
-\,{dE\over dx}\equiv  R(E)\simeq 
{\rho\over\rho_W}\,R_0\,(1+B\,E)\; ,
\label{loss1}
\end{equation}
where $\rho$ is the material's density, $\rm\rho_W$ is 1 g cm$^{-2}$, and
\begin{equation}
\rm R_0\simeq 2.12\;\left[{2\,Z\over A}\right]\;{MeV\over cm};\;\;\;\;\;
 B\simeq 0.125\;{Z\over TeV}\; .
\label{loss2}
\end{equation}
In ice or a typical rock material $\rm 2\,Z/A\simeq 1$, and 
$\rm Z$ is small enough for the neglect of the B term in Eq.(\ref{loss1}), 
at the muon energies we shall encounter ($\rm E_\mu \ll 1$ TeV), 
 to be a good approximation.

At a given position x in a target material, an (approximately x-independent)
$\nu_\mu$ flux per unit area $\rm dN_\nu(E_\nu)/dE_\nu\,dA$ gives
rise to a $\mu$ flux $\rm dN_\mu(E_\mu,E_\nu,x)/dE_\nu\,dE_\nu\,dA$
satisfying the equation:
\begin{equation}
\rm
{\partial\over\partial x} \left[{dN_\mu\over dE_\nu\,dE_\nu\,dA}\right]=
\rho\,N_A\;{d\sigma\over dE_\mu}\;
{dN_\nu\over dE_\nu\,dA}
+{dE\over dx}\;{\partial\over\partial E_\mu}
\left[{dN_\mu\over dE_\nu\,dE_\nu\,dA}\right]\; .
\label{muonx}
\end{equation}
For a target thickness much larger than the muon range, an equilibrium
between the produced and slowed-down muons is reached, whereby the muon
flux is independent of position and the l.h.s.~of Eq.(\ref{muonx}) vanishes.
Inserting Eqs.(\ref{loss1}) and (\ref{loss2}) into 
Eq.(\ref{muonx}) and integrating, we obtain \cite{DeR}: 
\begin{eqnarray}
\rm
{dN_\mu\over dE_\mu\,dA}&=& \rm
\int_{E_\mu} K\;dE_\nu\,{dN_\nu\over dE_\nu\,dA}\,
(E_\nu-E_\mu)\; , \nonumber\\
\rm K &\simeq & \rm 
\rho_W\,N_A\,{1\over R_0}
\,{\sigma_{_{CC}}\over E_\nu}\simeq 2.26 \times 10^{-12}\;GeV^{-2}\; .
 \label{fluxmu1}
\end{eqnarray}

Define $\rm x_\mu=E_\mu/E_p$: the ratio of the energy of a muon produced on 
Earth to the energy $\rm E_p=m_p\,c^2\,\gamma_{in}$ of the CB's nucleons,
as they enter the SNS.
Substitute the neutrino flux of Eq.(\ref{nunum}) into Eq.(\ref{fluxmu1})
and integrate over neutrino energies to obtain a muon flux
per incident CB:
\begin{eqnarray}
\rm {dn_\mu \over dx_\mu\, d\Omega}&\sim&\rm K\,E_p^2\; 
\int_{x_\mu} {dn_\nu \over dx_\nu\, d\Omega}\,(x_\nu-x_\mu)\,dx_\nu\nonumber\\
&=&\rm K\,E_p^2\,{N_b\,B_\nu\over 2\,\pi\, D_L^2}\;(1+z)^4\;
f_\mu(z,\gamma_{in},\theta,x_\nu)\nonumber\\ \rm
f_\mu&=&\rm 
\;{2+c_\nu\,x_\mu\over c_\nu^3}\;e^{-c_\nu\,x_\mu}\, ,
\label{muhere}
\end{eqnarray}
with $\rm B_\nu$ and $\rm c_\nu$ as in Eq.(\ref{xtheta3})
and $\rm N_b$ the total baryon number of the CB, Eq.(\ref{NB}).
In Fig.(\ref{spectrax}c)  we show $\rm f_\mu$ as a function of
$\rm x_\mu$ at various $\theta$,
for $\rm z=1$ and $\rm\gamma_{in}=10^4$.

Very roughly characterize the efficiency of an experiment as
a step function jumping from zero to unity at $\rm E^\mu=E^\mu_{min}$.
The observable number of muons per CB and
per unit area, obtained by integration of Eq.(\ref{muhere}), then is: 
\begin{eqnarray}
\rm {dn^T_\mu[x^\mu_{min},\theta]\over d\Omega}&\sim&\rm 
{dn^T_\mu[0,0]\over d\Omega}\;G_\mu(z,\gamma_{in},\theta,x^\mu_{min})
\nonumber \\ \rm
G_\mu&=&\rm \left[{(1+z)\,b_\nu\over c_\nu}\right]^4
\;\left(1+{c_\nu\,x^\mu_{min}\over 3}\right)\;e^{-c_\nu\,x^\mu_{min}}\nonumber\\
\rm x^\mu_{min}&\equiv&\rm {E^\mu_{min}\over m_p \,\gamma_{in}}
\nonumber\\ \rm
{dn^T_\mu[0,0]\over d\Omega}
&\simeq&\rm {3.2\times 10^2\over km^2}\,{\overline E_{CB}}\,
\left[{\gamma_{in}\over 10^4}\right]^3\,
\left[{D_L(1)\over D_L(z)}\right]^2\, .
\label{muonseen}
\end{eqnarray}

In Figs.(\ref{Gmuon}a,b) we show $\rm G_\mu$ as a 
function of $\rm x^\mu_{min}$ at various fixed $\theta$, and vice versa;
for $\rm z=1$ and $\rm\gamma_{in}=10^4$. The relatively large flux
$\rm dn^T_\mu[0,0]/ d\Omega$ of Eq.(\ref{muonseen}) is seen
to be very significantly reduced as $\theta$ and/or $\rm x^\mu_{min}$
 depart from zero. Once again, for
a GRB with $\rm n_{CB}$ significant pulses,
the total number of muons is $\rm n_{CB}$ times
larger than that of  Eq.(\ref{muonseen}), and neutrino oscillations
may reduce the $\nu_\mu$ flux by a factor 2 or 3.

\section{Angular apertures and observational prospects}

Barring the case of GRB 980425 ---whose exceptional
properties and their interpretation within the CB model
are discussed in \cite{DD2000b}--- the equivalent spherical energies 
of the GRBs with measured redshifts range between 
$\sim 2 \times 10^{54}$ erg (GRB 990123) and $\sim 2 \times 10^{51}$ erg 
 (GRB 970228). The dependence of the GRB flux on the angle $\theta$
subtended by the CB's velocity vector and the line of sight
is given by Eq.(\ref{dfdomega}): 
$\rm df/d\Omega\propto (1+\theta^2\,\gamma_{out}^2)^{-3}$.
This $\theta$ dependence is the steepest parameter
dependence of the CB model, see Fig.1 of \cite{DD2000b}.
It is therefore reasonable to attribute the range of observed equivalent
spherical energies to the $\theta$ dependence, as if GRBs were
otherwise approximately standard candles. The observed three orders of
magnitude spread in equivalent energy then corresponds, according
to Eq.(\ref{dfdomega}),  to a spread of viewing angles between 
$\theta \approx 0$ and $\rm\theta\approx 3/\gamma_{out}$. The cutoff at
the upper angle reflects the sensitivity of past and current observations.

The energies of the individual GRB $\gamma$-rays and the
GRB fluences indicate CB Lorentz factors $\rm\gamma_{out}\sim 10^3$.
So does an approximately 1:1 SN/GRB association. (The
geometrical fraction of currently observable GRBs, for 
$\theta < 3/\gamma$, is $\rm\pi \theta^2 / (4 \pi) \approx 9 / (4 \gamma^2)$.  
For $\gamma=10^3$ this fraction precisely reconciles the 
SN II, Ib, Ic rate in the observable universe: $12 \pm 5$ s$^{-1}$ \cite{Madau}
with the corresponding GRB rate of $\sim 10^3$ per year.)

The fluxes of $\nu$-induced muons and of $\gamma$'s
of GRB and EGR energies 
have different $\theta$ dependences and  the circumstance that GRBs are 
currently observed at angles up to 
$\rm\theta\sim 3/\gamma_{out}\sim 3\times10^{-3}$ plays an obvious role
in discussing the search for EGR and $\nu$ signals in spatial coincidence
with GRBs. The discussion is summarized
in Figs.(\ref{angdistrrs3}), where we compare the angular apertures of
the three fluxes. The absolute and relative normalizations in these
figures are arbitrary, so that the GRB results, based on
Eq.(\ref{dfdomega}), depend only on $\rm\gamma_{out}$,
chosen to be $10^3$. The EGR results, based on 
the second of Eqs.(\ref{photflux2}),
depend also on $\rm z$ (chosen at $\rm z=1$) 
and on $\rm E^\gamma_{min}$, taken here to be 50 GeV.
The $\nu$ results, also for $\rm z=1$, are based on the second
of Eqs.(\ref{muonseen}); they are for $\rm \gamma_{in}=10^4$ and
$\rm E^\mu_{min}=50$ GeV.

According to Figs.(\ref{angdistrrs3}), the EGR
beam, up to very large $\theta$, has a broader tail than the GRB beam.
In practice that means that a detector with the sensitivity to observe
the EGR flux of Eq.(\ref{photflux2}) should find a signal in temporal
and angular coincidence with a large fraction of detected GRBs.
The $\nu_\mu$-induced $\mu$ beam is about an order of magnitude
narrower than the GRB beam in angle, two orders of magnitude
in solid angle. Consequently, a detector with a sensitivity close to
that necessary to observe the $\mu$ flux of Eq.(\ref{muonseen})
would see coincidences with only about one in a hundred intense
GRB events, that is $\sim 1$\% of GRBs in
 the upper decade of observable fluences,
for which $\rm \theta\sim 1/\gamma_{out}$.

To ascertain the observational prospects for EGRs and $\nu$'s,
one would have to convolute our predicted fluxes
with the sensitivities of the many large-area or large-volume
$\nu$ and EGR ``telescopes''
currently planned, deployed or under construction.
We do not have sufficiently detailed information to do so,
but a coarse look at their potential indicates that
testing the CB model will neither be trivial, nor
out of the question. The small area of past detectors with
a capability to see EGRs, such as EGRET, would preclude
the observation of the flux of Eq.(\ref{photflux2}). 

\section{Timing considerations}
 
In the cannonball model, each CB crossing the SNS generates
an individual $\gamma$-ray pulse in a GRB light curve. The
complementary statement need not be true: not every
observed pulse necessarily corresponds to a single CB, since
the $\gamma$ rays generated by sufficiently close CBs may
overlap. This can be seen in the two top entries in Fig.(\ref{lightcurves}),
which show the lightcurves of the same ensemble of CBs
crossing two SNSs, which differ only in their density-profile index;
in the case of the more extensive SNS ($\rm n=4$) the various
CBs blend into a single pulse.

Each CB should generate three distinct pulses: a GRB pulse,
a $\nu$ pulse and an EGR pulse. The $\nu$ and EGR pulses
are narrower in time than the GRB pulse and they preceed it.
Observed with neutrinos or EGRs, then, a burst has the same
pulse structure as the GRB, but the pulses are shorter and
are precursors of the GRB pulses. We proceed to estimate
the magnitude of these effects, illustrated in Fig.(\ref{lightcurves}).

For a given density distribution of the SNS, such as that in
Eq.(\ref{profile}), it is possible, though laborious, to explicitly compute the
expected time profile of the neutrino signal. This profile is sensitive
to the shape of $\rm\rho(x)$ at all $\rm x$, including the inner part of the
shell, for which no empirical data are available.
Consequently, we shall only give here approximate results for the
width in time of the $\nu$ signal, and for its timing relative
to the onset of a GRB pulse.

Let $\rm x_\nu$ be the distance from the SN centre at which the
shell's grammage, as in Eq.(\ref{SNgram}), is half of the total
SNS grammage, that is $\rm x_\nu=R_s\,2^{1/(n-1)}$. The temporal
half-width of the neutrino signal, $\rm t_\nu$, is roughly the time
it takes the CB to reach this point\footnote{The time needed
for the ``last proton'' of the CB to catch up and interact with
the rest of the colliding CB is shorter  than $\rm t_\nu$ by a factor $\rm\sim\beta_{in}\,(\gamma_{out}/\gamma_{in})^2\,
R_S/(x_\nu-R_S)$.}. As measured
by the observer, this time is given by the same expression
as Eq.(\ref{ttp}) with the substitution of $\rm x^{tp}_{GRB}$ by
$\rm x_\nu$. The ratio of durations
of a single pulse in neutrinos and in GRB $\gamma$-rays is:
\begin{equation}
\rm
{t_\nu\over t_{GRB}}\sim {x_\nu-R_S\over x^{tp}_{GRB}-R_S}\, .
\label{nufrac}
\end{equation}
For SNS density indices $\rm n=8$, 6 and 4, this ratio is 0.038, 0.029
and 0.013, respectively: the duration of a $\nu$ pulse is a few per cent
of that of an individual GRB pulse.
Neutrinos are emitted from the moment the CB hits the SN shell,
while GRB $\gamma$ rays can only be seen if emitted from the
transparent SNS outer layer. The time difference between the onset
of the corresponding pulses is $\rm \sim t_{GRB}$: a $\nu$ pulse
should precede its corresponding GRB pulse by approximately
the width of the GRB pulse.

The discussion of the EGR pulse follows analogous lines. The ratio
of EGR and GRB pulse widths is:
\begin{equation}
\rm
{t_{EGR}\over t_{GRB}}\sim {x^{tp}_{EGR}-R_S\over x^{tp}_{GRB}-R_S}\, ,
\label{EGRfrac}
\end{equation}
with $\rm x^{tp}_{EGR}$ given by Eq.(\ref{xtp2}).
For SNS density indices $\rm n=8$, 6 and 4, this ratio is 0.48, 0.42
and 0.29, respectively: the duration of an EGR pulse is shorter than
that of the corresponding GRB pulse by a factor 2 or 3.
The time difference between the onset
of the corresponding pulses is a fraction  $\rm 1-{t_{EGR}/ t_{GRB}}$
of the duration of the GRB pulse: an EGR pulse
should preceed its corresponding GRB pulse by 50 to
70\% of the width of the GRB pulse. 

The light curve of an EGR pulse is proportional to the SNS shell
density, corrected for absorption. For the density profile of Eq.(\ref{profile})
and the corresponding grammage of Eq.(\ref{SNgram}):
\begin{eqnarray}
\rm
{dN_{EGR}\over dt}&\propto& \rm \rho_{_S}(x[t])\;
Exp \left[- \, {X_S (x[t]) \over X_{EGR}} \right] 
\nonumber\\ \rm
x[t]&=&\rm {\gamma_{out}\,\delta\over 1+z}\;c\,t\, .
\label{EGRdNdt}
\end{eqnarray}

The considerations of this section are visualized in Fig.(\ref{lightcurves}),
where we have drawn the light curves of a single GRB in 
GRB $\gamma$-rays, in EGRs and in neutrinos. The timing sequence
of the pulses is put in by hand and their normalizations correspond
to (random) values of $\rm\gamma_{out}$ close to $10^3$, see 
Eq.(\ref{newenergy}). The two columns of the figure correspond
to $\rm n=8$ and $\rm n=4$. Notice how the EGR pulses precede
the GRB pulses and are narrower: the EGR has a better time
``resolution''. For neutrinos, this is even more so.

\section{Conclusions}

\begin{figure}
\begin{center}
\vspace*{1.0cm}
\hspace*{-1cm}
\epsfig{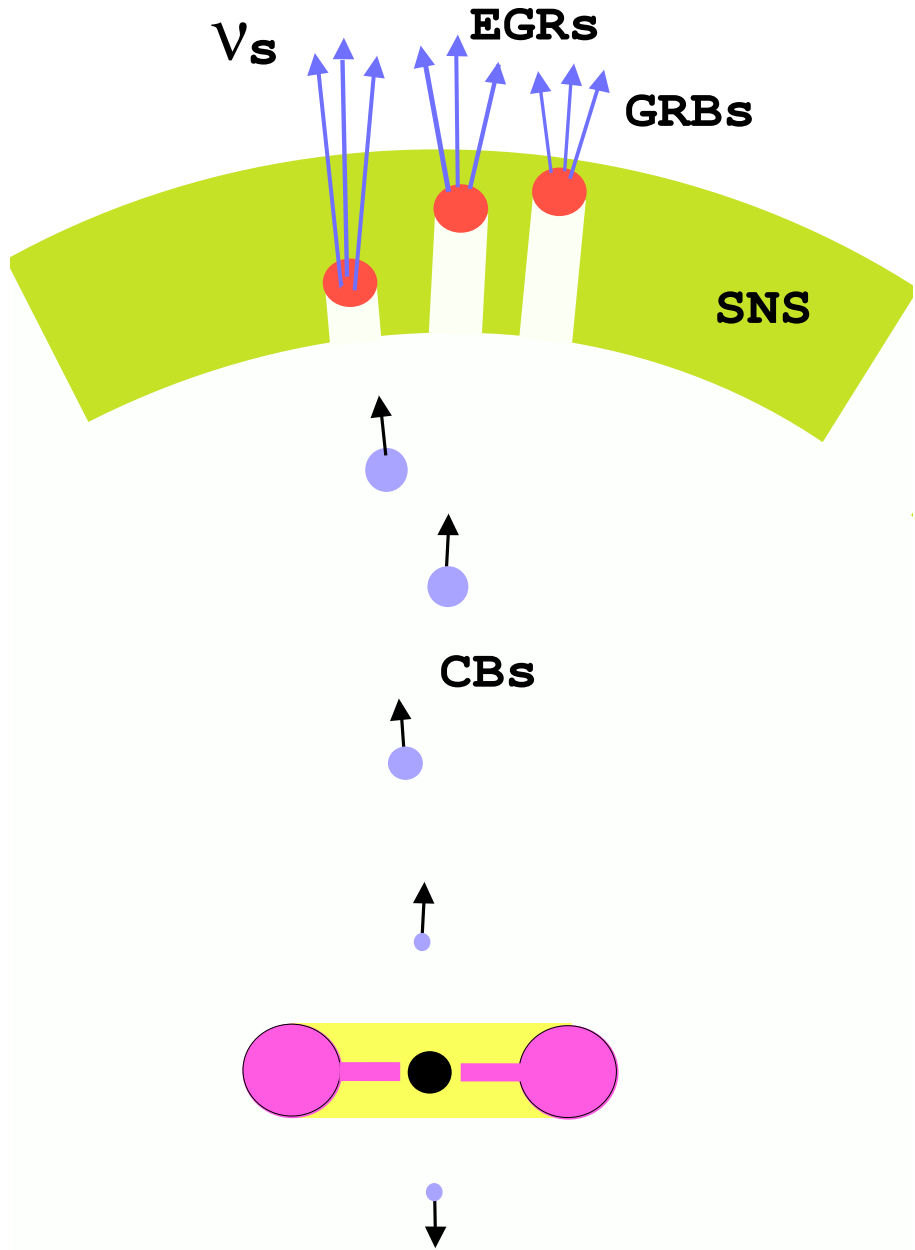}
\caption{The CB model in a SN environment, not shown to scale.
Relativistic CBs are emitted by a compact object
accreting matter from a disk and/or torus.
They hit a SN shell generating $\nu$'s,
quasi-thermal radiation
(the GRB) and $\gamma$-rays from $\pi^0$ decay
(the EGRs). The latter two exit only from the
transparent outer layers of the SN shell.}
\vspace*{-0.5cm}
\label{model}
\end{center}  
\end{figure}

\begin{figure}
\begin{center}
\vspace*{1.0cm}
\hspace*{-1cm}
\epsfig{file=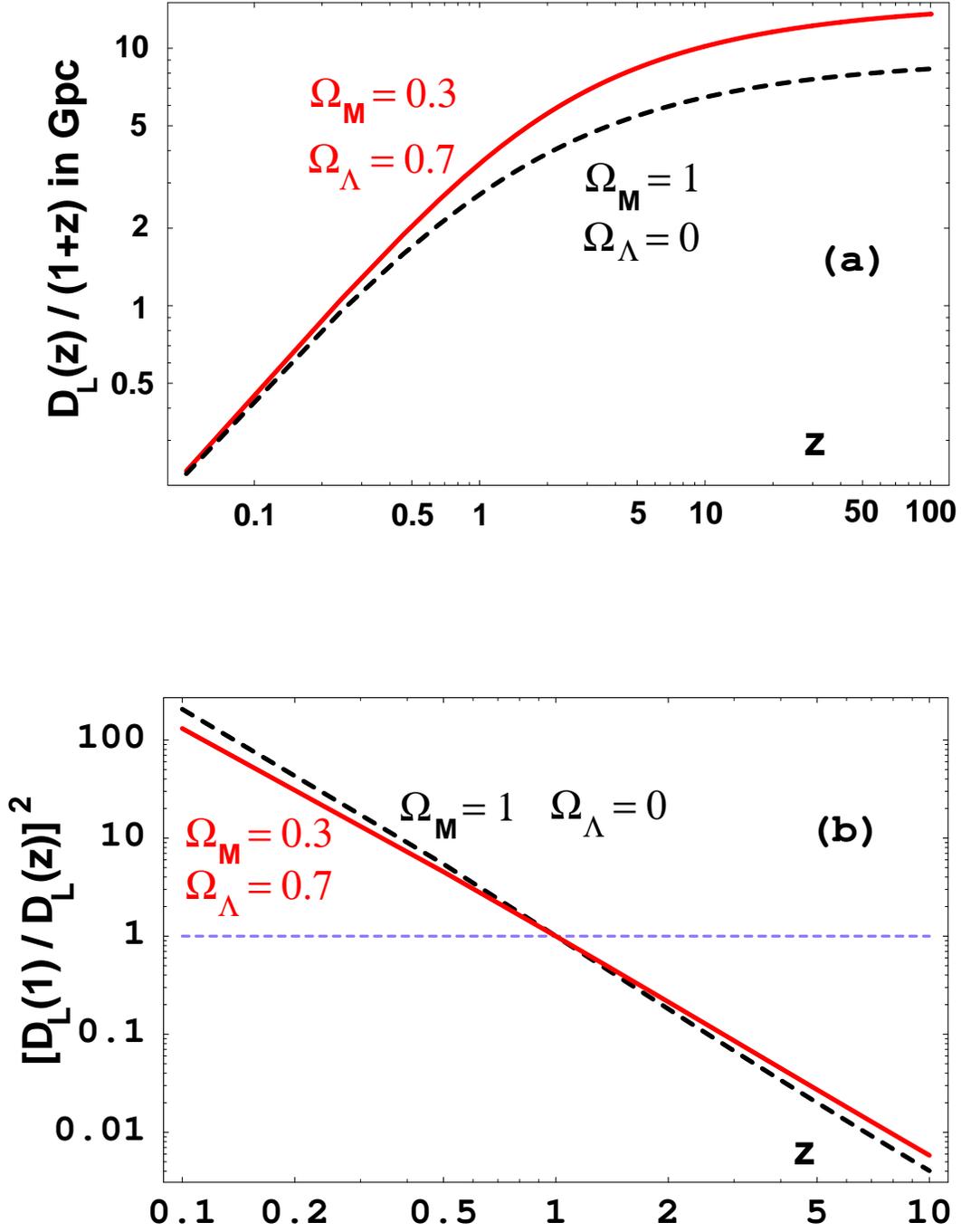,width=14cm}
\caption{Luminosity distances and ratios thereof, as functions of redshift,
for two $\Omega=1$ Friedman universes, with
two choices of matter and vacuum densities.}
\vspace*{-0.5cm}
\label{lumdis}
\end{center}  
\end{figure}

In the CB model of GRBs, illustrated in Fig.(\ref{model}), cannonballs heated
by a collision with intervening material produce GRBs by
thermal emission, and their electron constituency
generates GRB afterglows by bremsstrahlung, synchrotron radiation
and inverse Compton up-scattering of these photons 
and the cosmic background radiation.
The material CBs hit is an excellent ``beam-dump'',
so that nucleon--nucleon collisions
generate a very intense and collimated flux of neutrinos.
Because of absorption, the emission  of energetic
$\gamma$-rays via $\pi^0$ production and decay
is much less efficient, but by no means negligible.

\begin{figure}
\begin{center}
\vspace*{1.0cm}
\hspace*{-1cm}
\epsfig{file=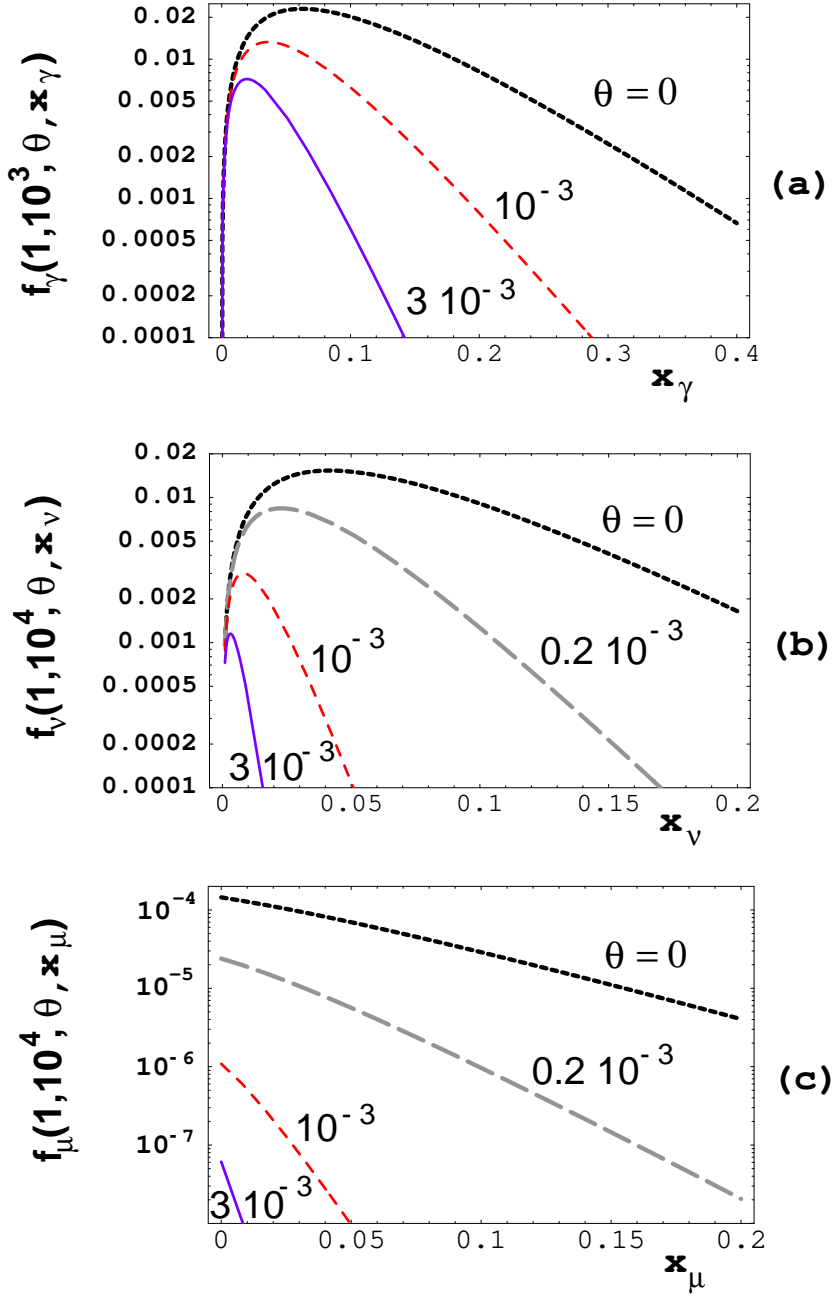,width=11 cm}
\caption{EGR, neutrino and muon fluxes, at various fixed observation
angles $\theta$, as functions of the fractional momentum of the
observed particle, at redshift unity. The functions 
$\rm f_\gamma(z,\gamma_{out},\theta,x_\gamma)$ of Eq.(\ref{photnum}),
for $\rm\gamma_{out}=10^3$, and 
$\rm f_{\nu,\mu}(z,\gamma_{out},\theta,x_{\nu,\mu})$ of 
Eqs.(\ref{nunum}, \ref{muhere}), both
for $\rm\gamma_{in}=10^4$, are depicted.}
\vspace*{-0.5cm}
\label{spectrax}
\end{center}  
\end{figure}

\begin{figure}
\begin{center}
\vspace*{1.0cm}
\hspace*{-1cm}
\epsfig{file=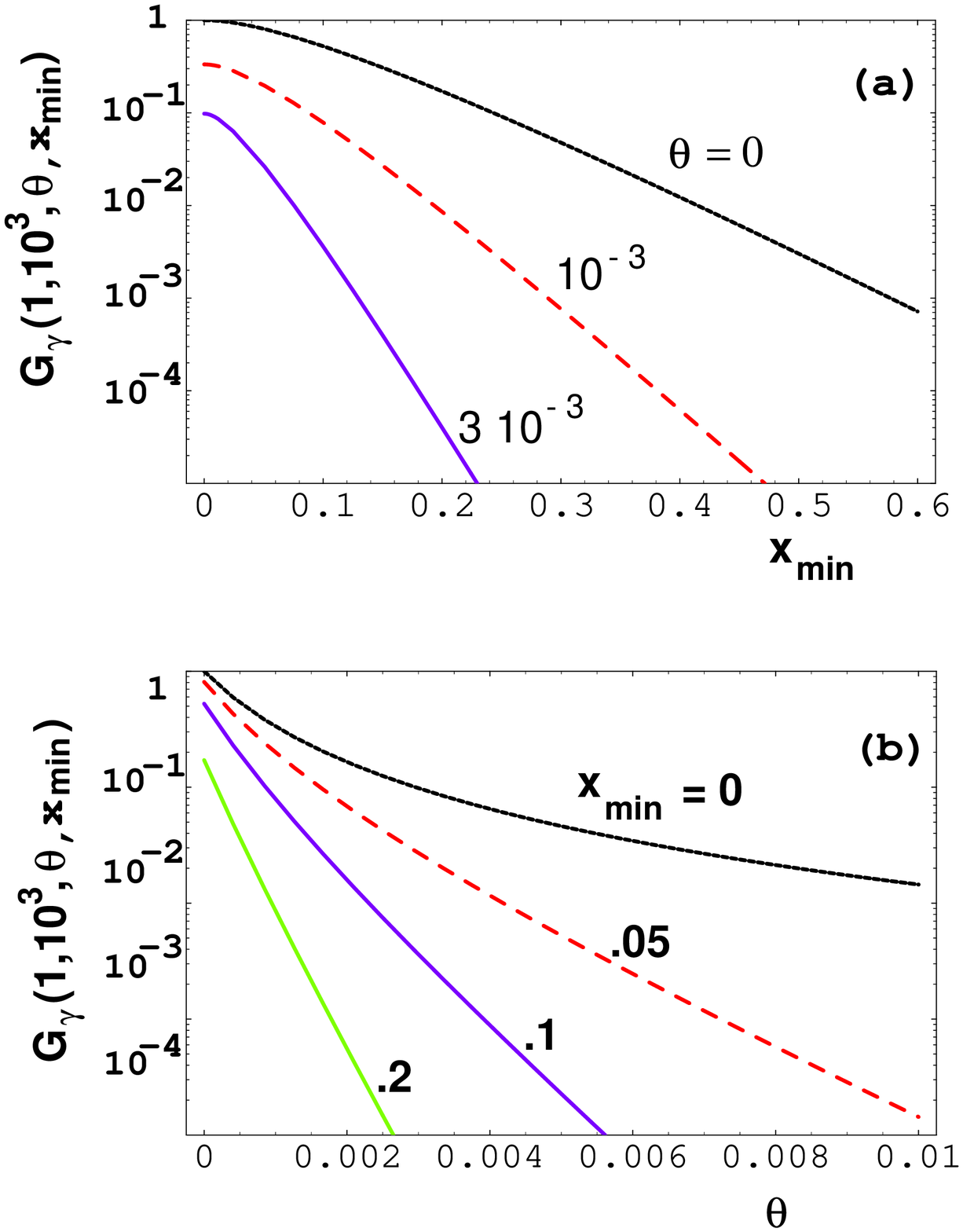,width=14cm}
\caption{The function $\rm G_\gamma$ of Eq.(\ref{photflux2}), for
$\rm z=1$ and $\rm \gamma_{out}=10^3$. Top: As a function of
$\rm x_{min}$ at various fixed $\theta$. Bottom: vice versa.}
\vspace*{-0.5cm}
\label{Ggamma}
\end{center}  
\end{figure}

\begin{figure}
\begin{center}
\vspace*{1.0cm}
\hspace*{-1cm}
\epsfig{file=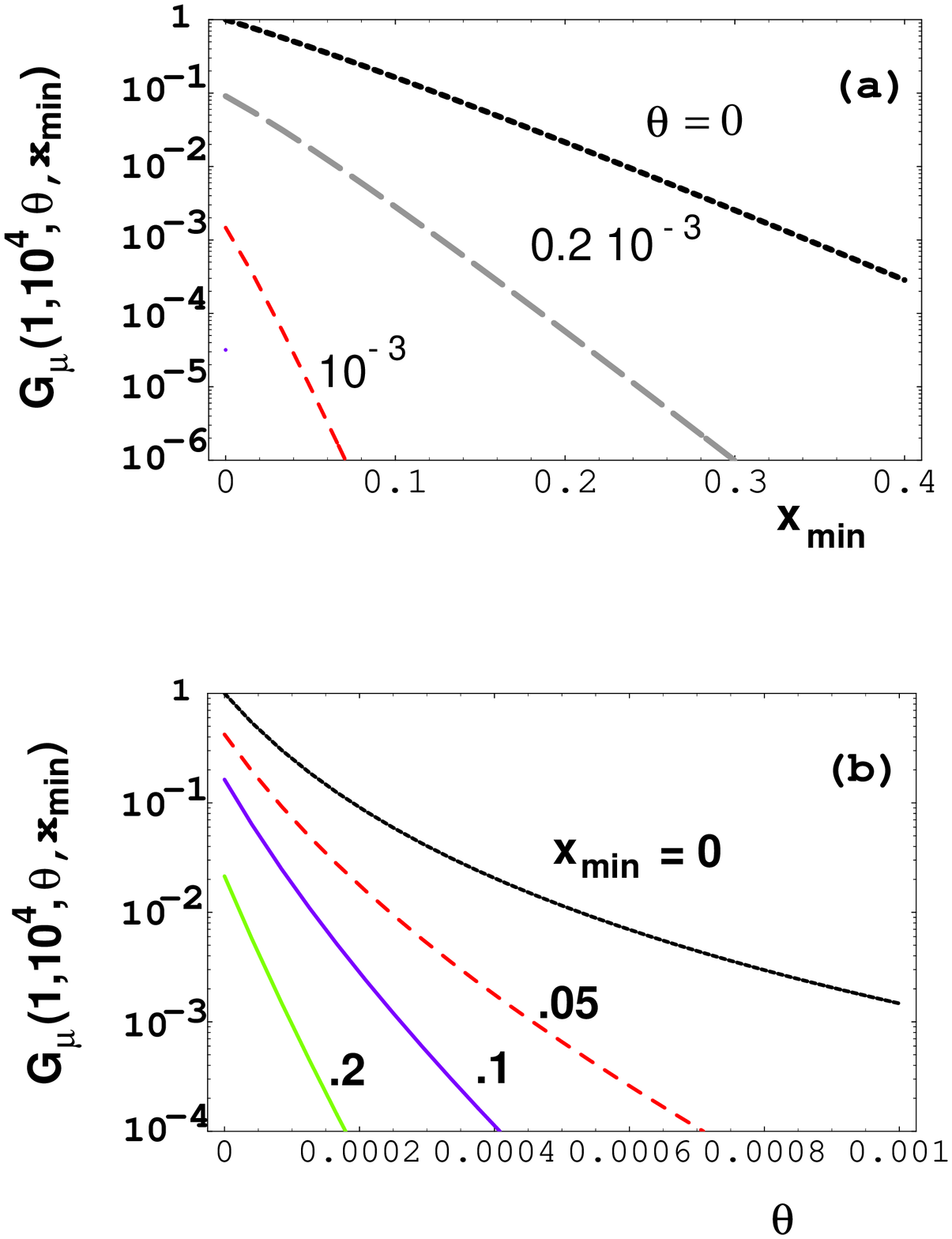,width=14cm}
\caption{The function $\rm G_\mu$ of Eq.(\ref{muonseen}), for
$\rm z=1$ and $\rm \gamma_{in}=10^4$. Top: As a function of
$\rm x_{min}$ at various fixed $\theta$. Bottom: vice versa.}
\vspace*{-0.5cm}
\label{Gmuon}
\end{center}  
\end{figure}

\begin{figure}
\begin{center}
\vspace*{1.0cm}
\hspace*{-1cm}
\epsfig{file=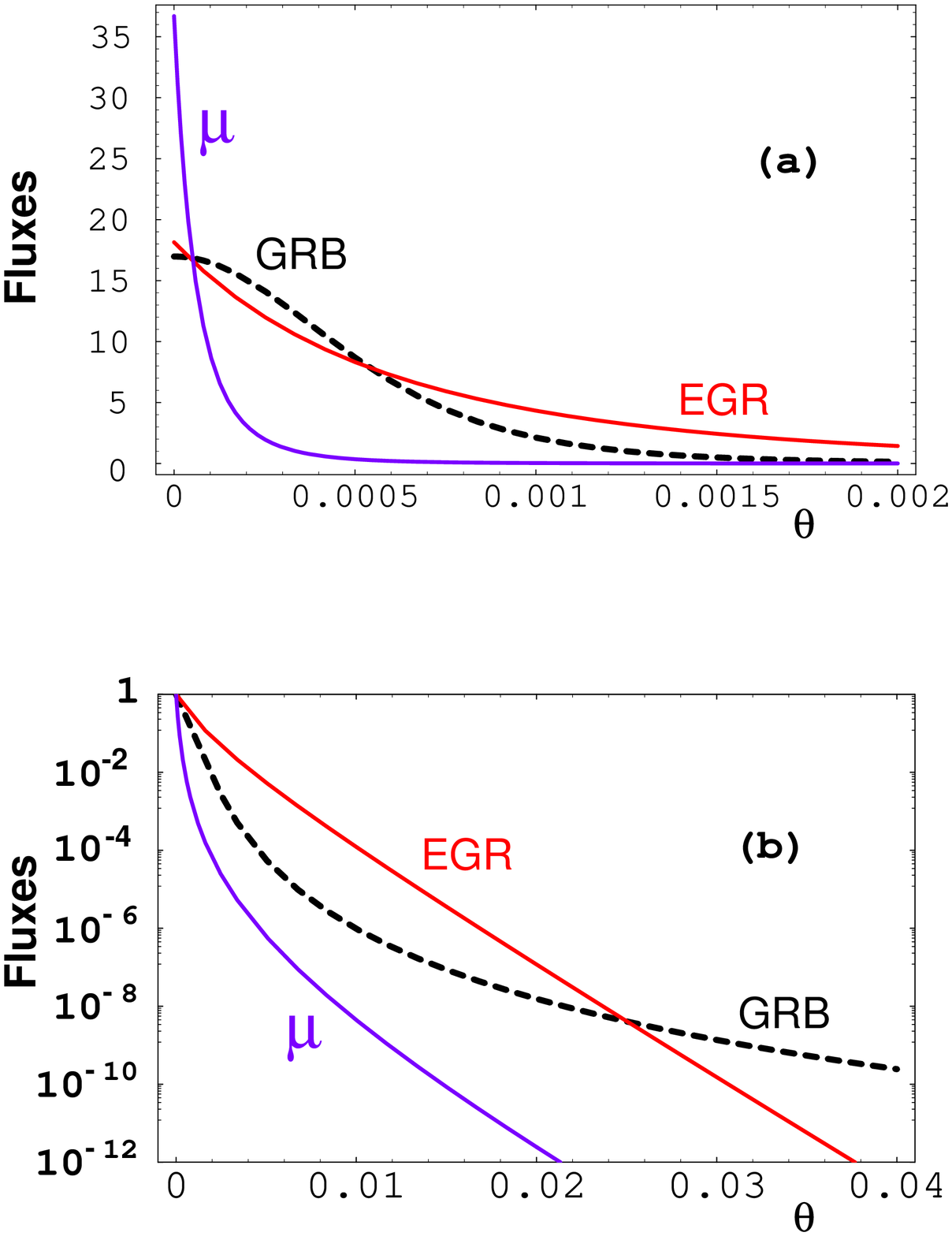,width=14cm}
\caption{Comparisons of angular distributions of
GRB photons, EGR photons and $\nu$-produced muons
in water or ice. In the upper graph, the normalizations
of the three curves are arbitrary. In the lower one,
they are all normalized to unity at $\theta=0$.}
\vspace*{-0.5cm}
\label{angdistrrs3}
\end{center}  
\end{figure}

The $\nu$ flux has a total energy of the order 
of $10^{53}$ erg (roughly 1/3 of the total energy in a jet
of CBs, augmented by the ratio 
$\rm \gamma_{out}/\gamma_{in}$, and reduced by the redshift factor).
But individual neutrinos have energies of only a few hundred GeV,
as illustrated in Fig.(\ref{spectrax}),
and their enormous flux will be hard to detect, even though it is
collimated within an angle $\sim 10^{-4}$. The detection in
coincidence with GRBs will be further hampered by the fact
that the GRB angular distribution is broader, as shown in
Fig.(\ref{angdistrrs3}).

The EGR flux carries roughly as much energy as the
GRB, that is $\rm E^{rest}_{pulse}\,\gamma_{out}\sim 10^{48}$ erg per pulse,
with $\rm E^{rest}_{pulse}$ as in Eq.(\ref{dfdomega}).
The EGR beam, as shown in Fig.(\ref{angdistrrs3}),
is somewhat broader than the GRB beam, so
that the search for coincidences should be fruitful. The typical
energies of EGRs, as illustrated in Fig.(\ref{spectrax}),
 are of tens of GeVs, and the relatively high threshold
energies of current large-area detectors should be a limiting issue,
as in the case of neutrinos.

\begin{figure}
\begin{center}
\vspace*{1.0cm}
\hspace*{-1cm}
\epsfig{file=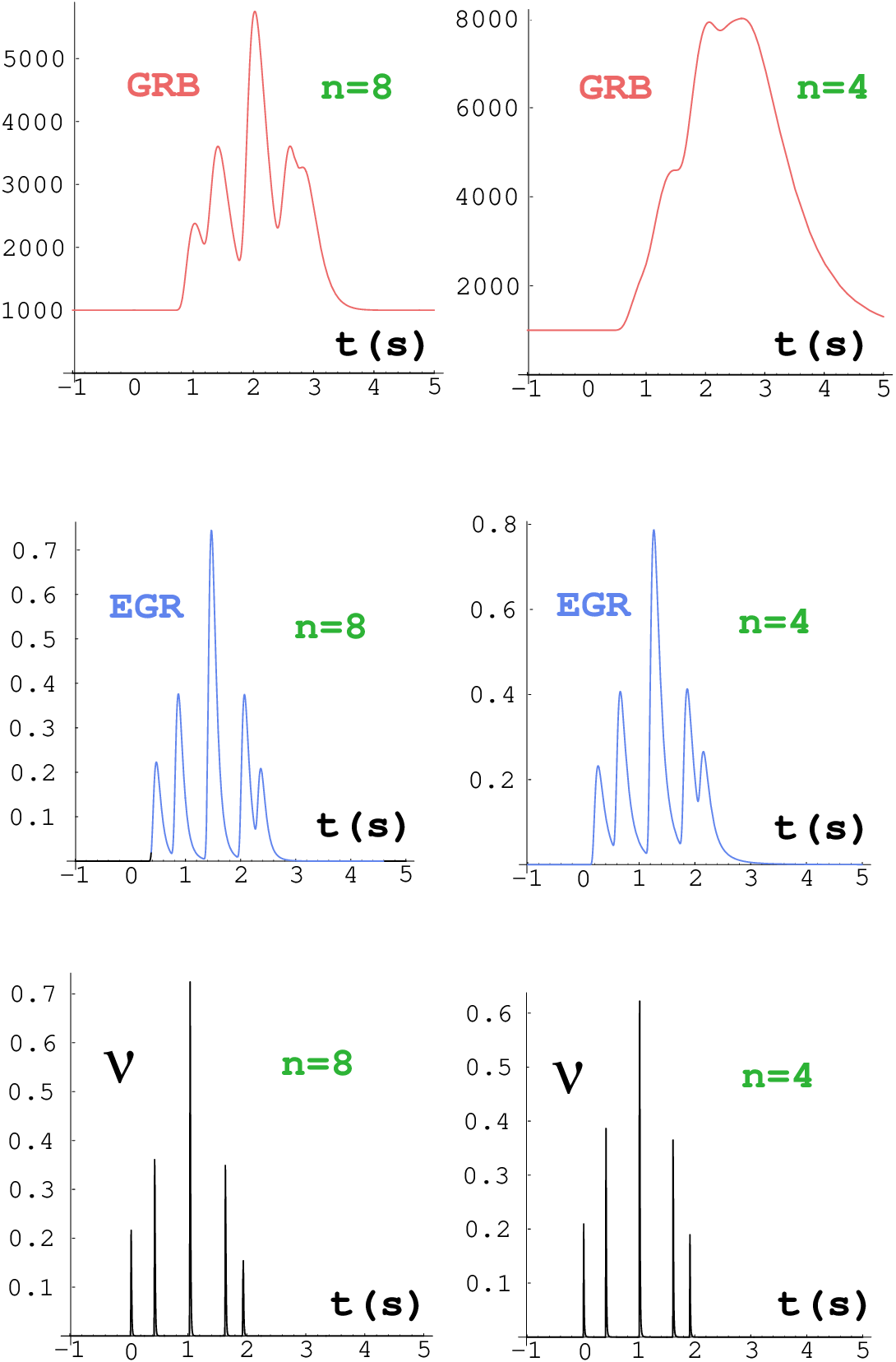,width=11.6cm}
\caption{A ``synthetic'' $\gamma$-ray burst consisting
of five CBs with $\rm\gamma_{out}$ within a factor
of 2 of $\rm\gamma_{out}=10^3$, with other parameters
at their reference values. The CBs are fired at random times in a 
2.5 s interval. The two columns are for SNS 
density indices $\rm n=8$ and 4. Top: the event seen
in the 30 keV to 1 MeV GRB domain. Middle: seen
in EGRs from $\pi^0$ decay. Bottom: the neutrino signal.}
\vspace*{-0.5cm}
\label{lightcurves}
\end{center}  
\end{figure}

The pulses of the GRB $\gamma$-rays should be slightly
preceded by narrower pulses of EGRs and by much
narrower pulses of $\nu$'s, as illustrated in Fig.(\ref{lightcurves}).
The CB model, as we have seen, predicts very specific properties and
relations between the GRB, EGR and $\nu$ spectra and light curves.
In this respect, as in many others, the Cannonball
Model is exceptionally falsifiable.

\noindent
{\bf Acknowledgement:} This research was supported in part by the 
Asher Fund For Space Research and the Fund For Promotion of Research 
at the Technion.

{}

\begin{thebibliography}{9}

\bibitem{Costa} 
E. Costa et al., Nature {\bf 387} (1997) 783.
\bibitem{vP} 
J. van Paradijs et al.,  Nature {\bf 386} (1997) 686.
\bibitem{Galama} 
T.J. Galama et al., Nature {\bf 395} (1998) 670.
\bibitem{Metz} 
M.R. Metzger et al., Nature {\bf 387} (1997) 878.
\bibitem{PG} 
B. Paczynski, Astroph. J. {\bf 308} (1986) L43; J. Goodman, 
Astroph. J. {\bf 308} (1986) L47;
J. Goodman, A. Dar and S. Nussinov, Astroph. J. {\bf 314} (1987) L7.
\bibitem{WM} 
S.E. Woosley, Astroph. J. {\bf 405} (1993) 273;
S.E. Woosley and A.I. MacFadyen, Astron. and Astroph. {\bf 138} (1999) 499;
A.I. MacFadyen and S.E. Woosley, Astroph. J.  {\bf 524} (1999) 168.
\bibitem{Pacz} 
B. Paczynski, Astroph. J.  {\bf 494} (1998) L45.
\bibitem{AHH}
J. Alvarez-Muniz, F. Halzen and  D.W. Hooper,
Phys. Rev. {\bf D62} (2000) 093015, and references therein.
\bibitem{SD} 
N.J. Shaviv and A. Dar, Astroph. J.  {\bf 447} (1995) 863.
\bibitem{Dar98}  
A. Dar, Astroph. J.  {\bf 500} (1998) L93.
\bibitem{DP} 
A. Dar and R. Plaga, Astron. and Astroph. {\bf 340} (1999) 259.
\bibitem{DD2000a} 
A. Dar and A. De R\'ujula, astro-ph/0008474,
and references therein.
\bibitem{DDD} 
S. Dado, A. Dar, R. Plaga and A. De R\'ujula, {\it Optical Afterglows of GRBs
in the Cannonball Model}, to be published.
\bibitem{Darz} 
A. Dar, Gamma Ray Communication Network report
No. 346 (1999), gcncirc@lheawww.gsfc.nasa.gov.
\bibitem{Reic}
D. Reichart,  Astroph. J.  {\bf 521} (1999) L111.
\bibitem{Galamab}
T.J. Galama et al.,  Astroph. J.  {\bf 536} (2000) 185.
\bibitem{Sok} 
V. Sokolov et al., astro-ph/0102492, Astron. and Astroph, in press. 
\bibitem{Holl}
S. Holland et al., astro-ph/0103058,  Am. Astron. Soc. 197 (2000) 63.03.  
\bibitem{Hj}
J. Hjorth et al., Astroph. J.  {\bf 534} (2000) L147. 
\bibitem {Sahu}
K.C. Sahu et al.,  Astroph. J.  {\bf 540} (2000) 74. 
\bibitem{CT}
A.J. Castro-Tirado et al., astro-ph/0102077, submitted to Astron. and Astroph.  
\bibitem{DD2000b} 
A. Dar and A. De R\'ujula, astro-ph/0012227.
\bibitem{DD2001} 
A. Dar and A. De R\'ujula, astro-ph/0102115.
\bibitem{MR} 
I.F. Mirabel and L.F. Rodriguez, Nature {\bf 371} (1994) 46, Ann. Rev.
Astron. and Astroph. {\bf 37} (1999) 409, astro-ph/0007010;
L.F. Rodriguez and I.F. Mirabel, Astroph. J.  {\bf 511} (1999) 398.
\bibitem{Ghis} 
G. Ghisellini et al., Astroph. J.  {\bf 407} (1993) 65.
\bibitem{Ked} 
L. Kedziora-Chudczer et al., Advances in Space Research {\bf 26} (2000) 727. 
\bibitem{Marg} 
B. A. Margon, Ann. Rev. Astron. and Astroph. {\bf 22} (1984) 507.
\bibitem{Kot} 
T. Kotani et al., Pubs. of the Astron. Soc. of Japan
 {\bf 48} (1996) 619. 
\bibitem{LL} 
A.G. Lyne and D.R. Lorimer, Nature {\bf 369} (1994) 127.
\bibitem{Celo} 
A. Celotti at al., Monthly Notices of the Royal Astron.. 
Soc. {\bf 286} (1997) 415.
\bibitem{Ghi} 
G. Ghisellini, astro-ph/0012125.
\bibitem{Naka} 
see, for instance, T. Nakamura et al., astro-ph/0007010.
\bibitem{Groom} 
see, for instance, D.E. Groom et al.,
{\it Review of Particle Physics}, Eur. Phys. J. {\bf C15} (2000) 1. 
\bibitem{FF} 
R. Fusco-Fumiano et al., Astroph. J. {\bf 513} (1999) L21;
J.S. Kaastra et al., Astroph. J. {\bf 519} (1999) L119.
\bibitem{YF} 
Y. Fuzukawa et al., astro-ph/00011257
\bibitem{SNR} For Cas A, see, for instance,
G.E. Allen et al., Astroph. J. {\bf 487} (1997) L97; 
F. Favata et al., Astron. and Astroph. {\bf 324} (1997) L49;
L.S. The et al., Astron. and Astroph. Supplement {\bf 120} (1996) 357.  
For IC 443, see, for instance, J.W. Keohane  et al.,
Astroph. J. {\bf 484} (1997) 350.
For RCW 86, see G.E. Allen et al.,  American Astron. Soc.
{\bf 193} (1998) 51.01.
\bibitem{Bailly} 
J.L. Bailly et al., Z. Phys. {\bf C35} (1987) 309.
\bibitem{Neu} For a review, see, for instance,
G. Giacomelli, Int. J. Mod. Phys. {\bf A5} (1990) 223.
\bibitem{BaBre} 
D.S. Barton et al., Phys. Rev. {\bf D27} (1983) 2580;
A.E. Brenner et al., Phys. Rev. {\bf D26} (1982) 1497.
\bibitem{DeR} 
A. De R\'ujula et al., Phys. Rep. {\bf 99} (1983) 341.
\bibitem{Madau} 
P. Madau, M. Della Valle and N. Panagia, 
Month. Not. Roy. Astron. Soc. {\bf 297} (1998) L17.
\end{thebibliography}
\end{document}